\documentclass{article}
\usepackage{slashed,verbatim,subfigure}
\usepackage[numbers,sort&compress]{natbib}
\usepackage{amsmath}
\DeclareUnicodeCharacter{00A0}{ } 
\usepackage{amssymb}
\usepackage{appendix}
\usepackage{graphicx}
\usepackage[T1]{fontenc}
\usepackage[utf8]{inputenc}
\usepackage{dcolumn}
\usepackage{bm}
\usepackage[colorlinks=true,linktocpage=true,
linkcolor=blue,citecolor=blue]{hyperref}
\usepackage[a4paper]{geometry}
\newcommand{\be}{\begin{eqnarray}}
\newcommand{\ee}{\end{eqnarray}}

\begin{document}
\large
\title{ \bf{Effect of color reconnection and hadronic re-scattering  on underlying events in pp collisions at LHC energies.}}
\author{Krishna ~Kumar\footnote{krishna19@iisertvm.ac.in}~~and~~Sadhana Dash\footnote{sadhana@phy.iitb.ac.in}\vspace{0.03in} \\
Indian Institute of Science Education and Research Thiruvananthapuram, Vithura 695551, India  \\
Department of Physics, Indian Institute of Technology Bombay, Mumbai 400076, India}

\date{}
\maketitle

\begin{abstract}

Underlying events dominate most of the hadronic activity in p$-$p collisions and are spanned from perturbative to non-perturbative QCD, having a sensitivity ranging from the multi-scale to very low-x scale physics. 
A detailed understanding of such events plays a crucial role in the accurate understanding of Standard Model ()SM and Beyond Standard Model physics. The underlying event activities has been studied within the framework of Pythia 8 Monte Carlo model, considering the underlying events observables mean charged particle multiplicity density , $\langle d^{2}N /d\eta d\phi \rangle$ and mean scalar $p_T$ sum,  $\langle d^{2} \sum p_{T} /d\eta d\phi \rangle$ as a function of leading charged particle in towards, away, and transverse region of p$-$p  collisions at $\sqrt{s}$  = 2.76, 7 and 13 TeV. The towards, away, and transverse regions have been defined on an azimuthal plane relative to leading particle in p$-$p collisions. The energy dependence of underlying events and their activities in the central and forward region has also been studied. The effect of hadronic re-scattering, color reconnection, and rope hadronization mechanism implemented in Pythia 8 has been studied in details to gain insight into the different processes contributing to underlying events in soft sector. 
\end{abstract}

\section{Introduction}
A quintessential proton-proton(p$-$p) collision can be sub-divided into two components: the primary hard partonic scattering and other associated activity collectively  termed as underlying events, UE. The cross-section of interactions involved in primary  hard-scattering processes can be determined using perturbative QCD (pQCD) calculations as they  involve sufficiently large momentum transfer  and, therefore the strong interaction coupling constant is small. On the contrary, the production mechanism of underlying events occur at a lower momentum scale. 
In order to understand the recent  Standard Model measurements at LHC and search for physical phenomena beyond Standard Model (BSM), it is crucial to have a detailed understanding of the underlying events in p$-$p collisions \cite{atlas1,field2012underlying}.\\
Experimentally, the underlying events can not be distinguished from the primary hard scattering process on an event by event basis, and one can not uniquely determine the origin of the final state hadrons. Therefore, the  observables generally chosen to study the underlying events receive contributions from both hard scattering and underlying events. However, one can utilize the topology of hadron-hadron collisions to understand the accompanying interactions in p$-$p collisions apart from the initial hard scattering one \cite{cms1,alice1}. 

\begin{figure}
\includegraphics[width=0.7\linewidth]{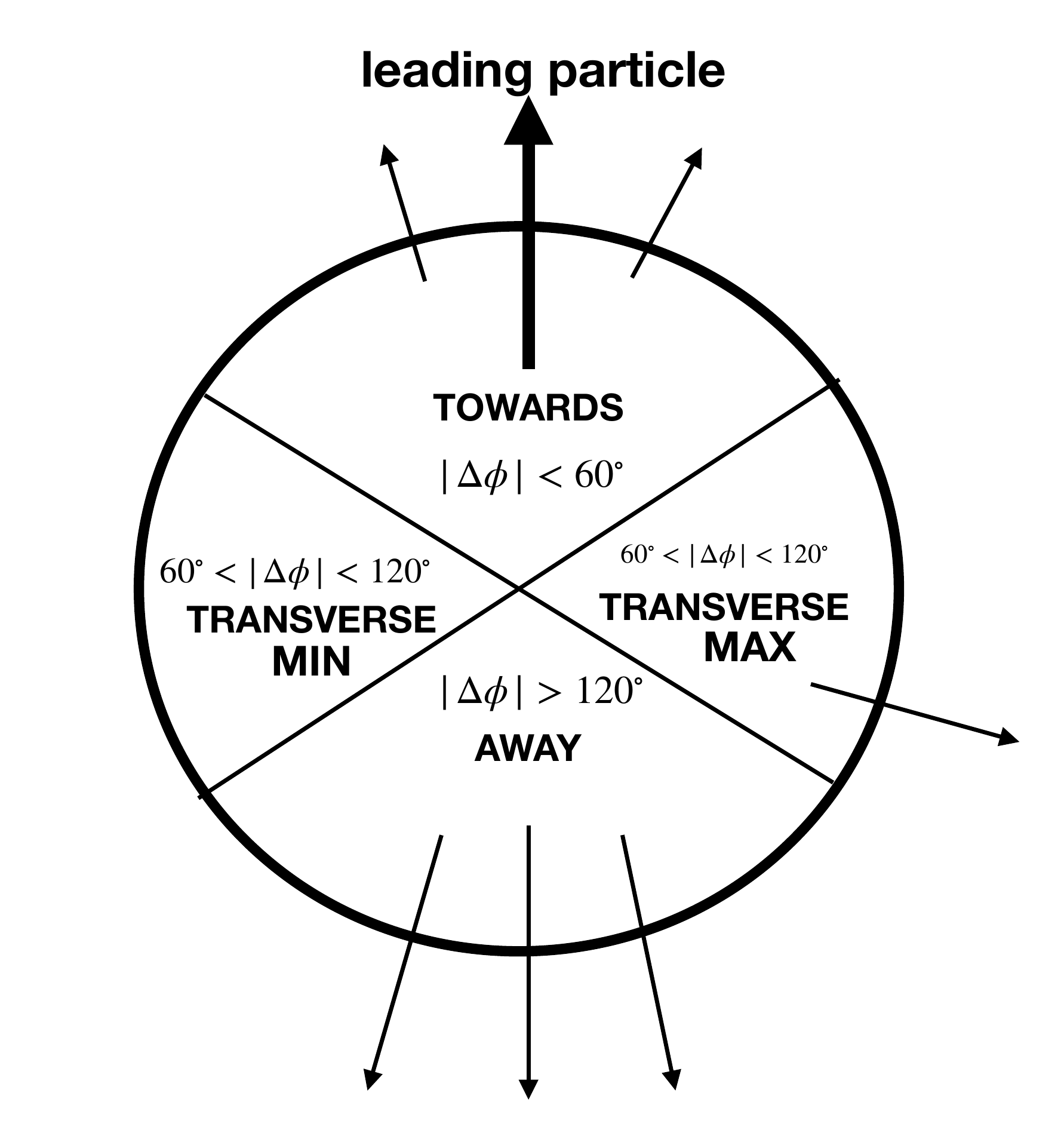}
\caption{The topology of  p$-$p collision where the azimuthal plane is segmented into towards, away and transverse regions in an event defined by $\Delta\Phi$ angle relative to leading particle direction
\cite{cms2015event,aaltonen2015study}.}
\label{fig1}
\end{figure}

The traditional approach had been used to study the observables sensitive to UE on an event by event basis. In this approach, the leading particle (particle with highest $p_{T}$) is used to segment 
the $\eta- \phi$ space into three distinct regions based on the azimuthal angular difference ($\Delta\Phi $) relative to the leading particle as shown in Figure \ref{fig1}. The leading particle acts as a proxy for the main flow of hard-scattering process \cite{atlas1}. The azimuthal angular difference is defined as, $\Delta\Phi = |\Phi-\Phi_L|$, where $\Phi$ is the azimuthal angle of an outgoing charged particle in an event and $\Phi_L$ is the azimuthal angle of the charged particle having highest transverse momentum ($p_T^{lead}$) in the event. The {\bf toward} region of $\eta-\Phi$ space is defined as $|\Delta\Phi| < 60^\circ$  and the {\bf away} region is defined as $| \Delta\Phi| > 120^\circ$. The two transverse regions  defined as  $60^\circ < \Delta\Phi < 120^\circ$ and $60^\circ < -\Delta\Phi < 120^\circ$  are referred to as transverse 1  and transverse 2  regions. The {\bf transverse} region is a combination of the transverse 1 and transverse 2  regions \cite{atlas1}. The towards and away regions are dominated by particles produced in hard processes, while the transverse region is more sensitive to underlying events.
In this work, an attempt has been made to study the underlying events through two observables, namely,  the mean charged particle multiplicity density, $\langle d^{2}N /d\eta d\phi \rangle$ and  
mean scalar $p_T$ sum,  $\langle d^{2} \sum p_{T} /d\eta d\phi \rangle$ in the three different regions for p$-$p collisions at  $\sqrt{s}$ = 2.76, 7 and 13 TeV. The energy dependence of underlying events  has been explored using Pythia 8 Monte Carlo model in the central region with $|\eta| < $ 2.5 and forward region with -6.6 $<$ $\eta$ $<$ -5.2 \cite{chatrchyan2013study}.  The effect of hadronic re-scattering, color reconnection, and rope hadronization mechanism have been explored to shed light on the underlying events. 

\section{PYTHIA 8 Monte Carlo Generator }

In this present work, Pythia 8.3 \cite{pythia8} has been used  for generating events, and ROOT \cite{root}  framework has been used to analyse the monte-carlo data. 
Pythia 8 is a Monte Carlo event generator for high-energy particle collisions to study collider physics. 
The hadronization framework is based on Lund string model.  It assumes that a string can be formed due to the colour field between two interacting quarks, and string potential energy linearly rises with the distance between two quarks. At some point, due to stretched string potential energy, it will break into hadrons. and the partons  emanate from the same vertex as one can anticipate for $e^{+}- e^{-}$ collisions. However, protons are composite objects and their overlap in hadronic collisions can lead to many partonic interactions in the region of overlap. This constitutes the MPI framework in Pythia. These fragmenting strings can overlap with each other  and the number of primary hadrons  produced  can be very high. In such a scenario, there is a high probability that the hadrons interact with each other ( elastically or inelastically) around the  region surrounding primary scattering. These hadronic re-scatterings can modify some of the event properties and therefore it will be worthwhile to study the effect of hadronic re-scattering on underlying event observables. The default parameters values given in Pythia 8.3 have been used in this  work for hadronic re-scattering model implementation \cite{pythiarescatter}.
Color reconnection is a microscopic mechanism that describes the interaction between colour fields. In this mechanism, the final partons are connected by colour strings 
in a way as to reduce the total string length to the minimum. One string that connects two partons follows their endpoint movements, resulting in a common boost of string fragmentation \cite{color1,color2}. 
Three different color reconnection mechanisms are implemented. The  MPI  based model connects 
 the partons of a lower $p_T$ MPI system  with one from a higher $p_T$ MPI system to reduce the total string length. In the QCD  based model, reconnection 
 happens if and only if it reduces the total string length and the string potential energy. One can form different quark junctions due to reconnection of hadronizing strings 
 iand QCD colour rules are used to determine the reconnection probability. The  Gluon move based model assumes that partons can move from one place to other  to reduce the total string length.\\
Rope hadronization model is an extension of the Lund string model which is significant when many overlapping strings are present. These strings  act coherently to form stronger ropes which would then be hadronized with larger, effective string tension. Due to rope formation, a smaller number of q$\overline{q}$ is needed to break the rope but having an effective string tension. There is reduction in the rope tension when new q$\overline{q}$ pair is produced in the process \cite{rope1,rope2}. Pythia 8 rope hadronization model describes the interaction between these strings by two mechanisms implemented as follows : \\
$\bullet$ String Shoving : The model allow nearby strings to shove each other with an interaction potential derived from the colour superconductor analogy. \\
$\bullet$ Flavor Ropes :The model assumes formation of ropes between strings overlapping in a dense environment which is hadronized with larger, effective string tension .\\

\section{Results and Discussion}

The observables used to study the underlying events are defined as follows :\\
$\bullet$ Mean charged particle multiplicity density, $\langle d^{2}N /d\eta d\phi \rangle$ : defined as the number of charged particles per unit pseudo-rapidity ($\Delta\eta$) per unit azimuthal angular difference ($\Delta\phi$).\\
$\bullet$ Mean charged particle scalar $p_T$ sum,  $\langle d^{2} \sum p_{T} /d\eta d\phi \rangle$ : defined as the $p_T$ sum of charged particles per unit pseudo-rapidity ($\delta\eta$) per unit azimuthal angular difference ($\delta\phi$) \cite{atlas1,atlas2}.\\
The transverse region is further subdivided into {\bf TransMAX} and {\bf TransMIN } region based on either transverse 1 or transverse 2 regions having the largest or smallest number of charged particles or mean scalar $p_T$ sum. The area factor is $\delta\phi = 2\pi/3$  for towards, away and transverse region, while for TransMAX and TransMIN region $\delta\phi = \pi/3$. The particles are selected to have $p_{T}$  $>$ 0.5 GeV/c, with $|\eta| \leq$ 2.5 .The underlying event activities in forward region have been explored for $-6.6 < \eta < -5.2$.

\begin{figure*}
\includegraphics[width=0.5\linewidth]{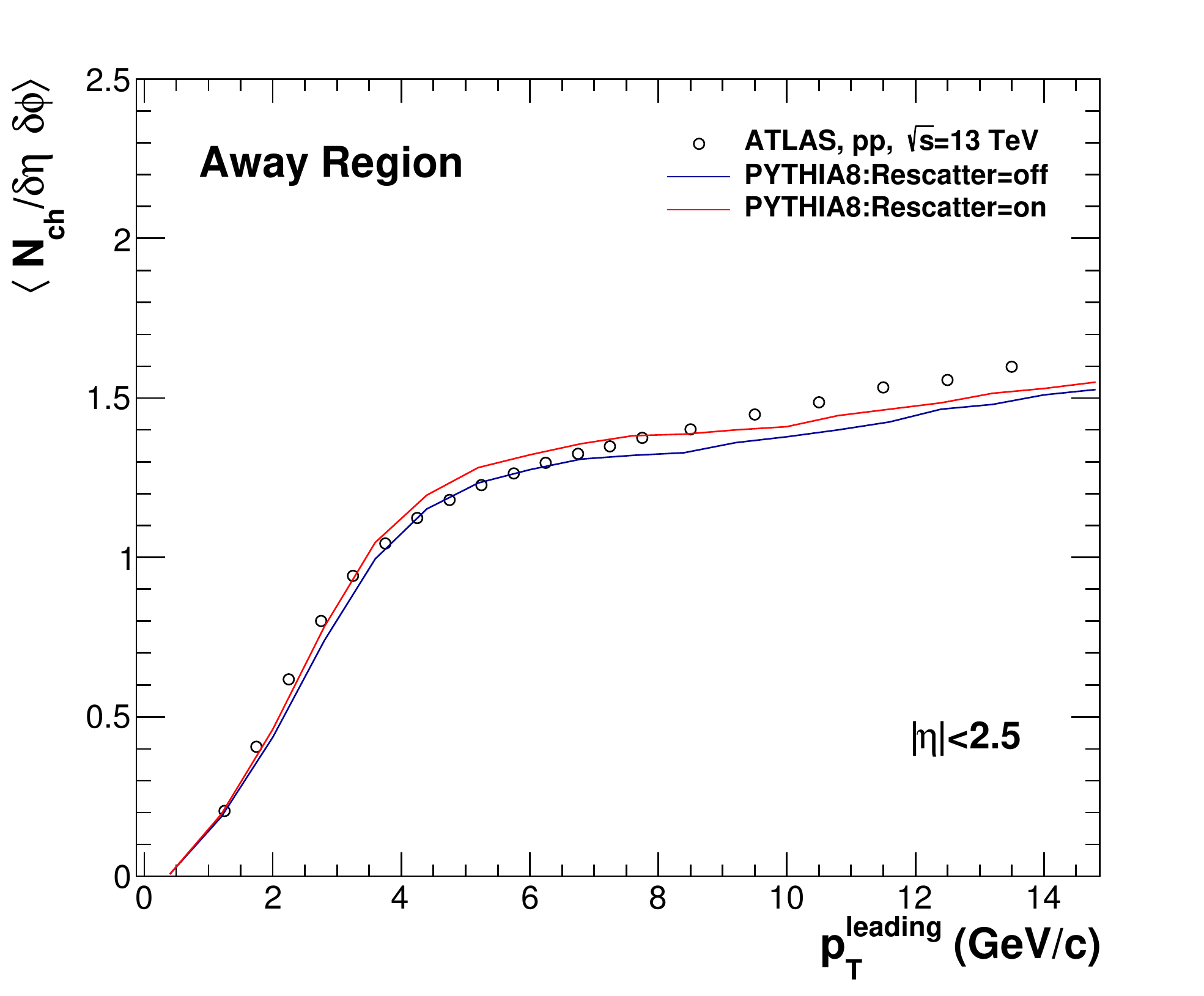}
\includegraphics[width=0.5\linewidth]{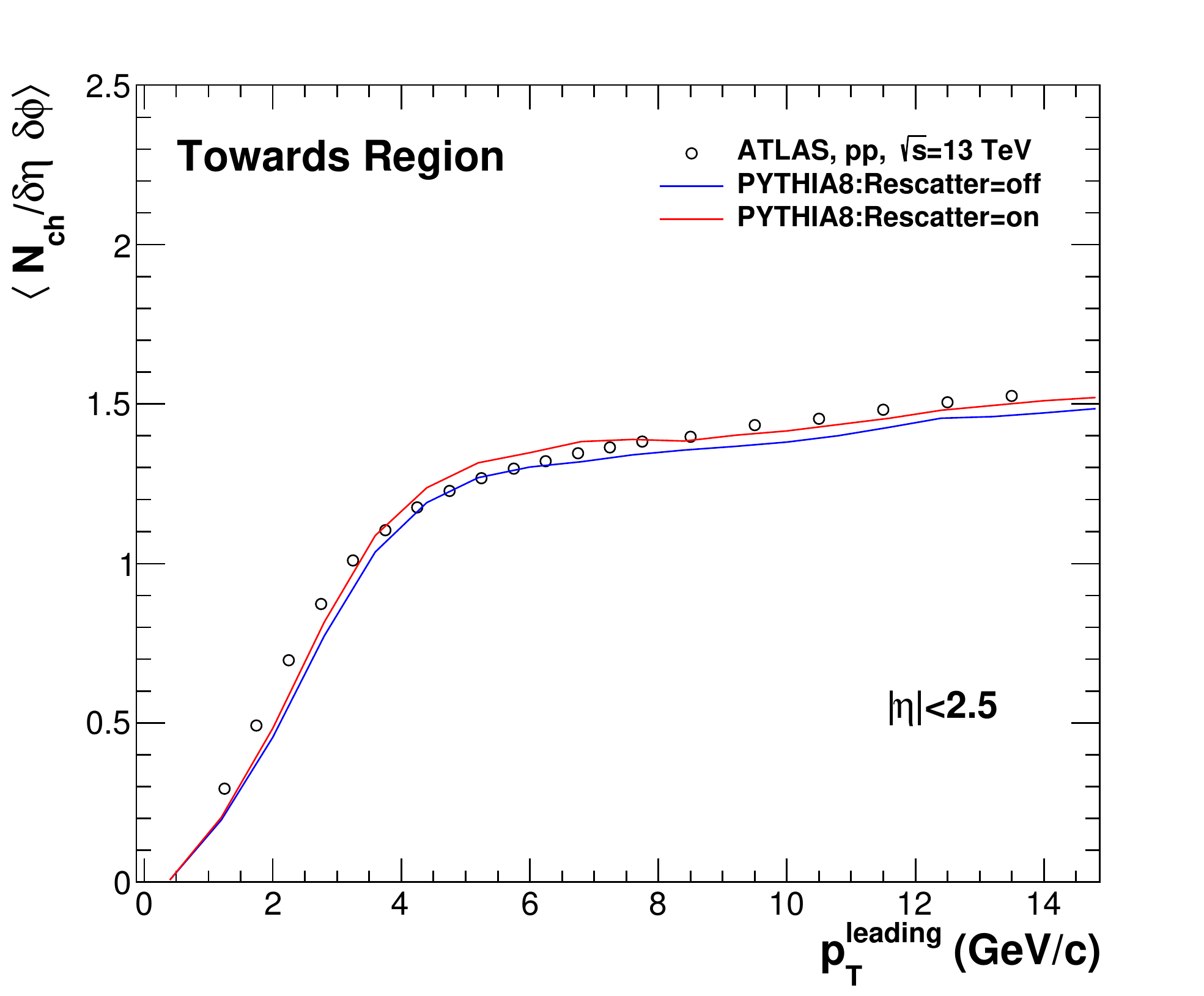}
\includegraphics[width=0.5\linewidth]{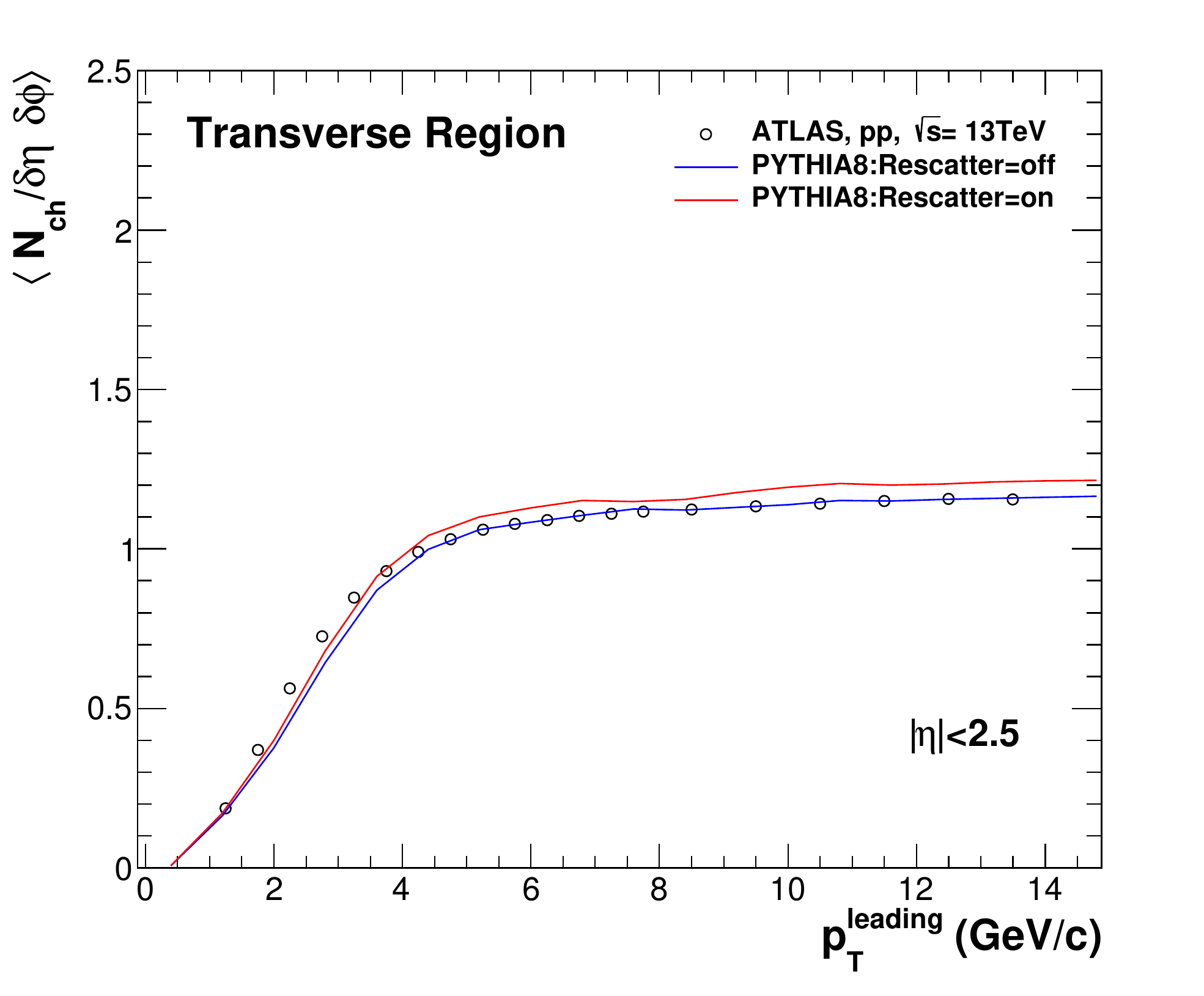}
\caption{$\langle d^{2}N /d\eta d\phi \rangle$ as a function of  $p_{T}^{lead}$  for towards, away, and transverse  regions in  p$-$p collisions at $\sqrt{s}$ = 13 TeV regions with(and without) the effect of hadronic re-scattering. }
\label{fig2}
\end{figure*}

\begin{figure}
\includegraphics[width=0.5\linewidth]{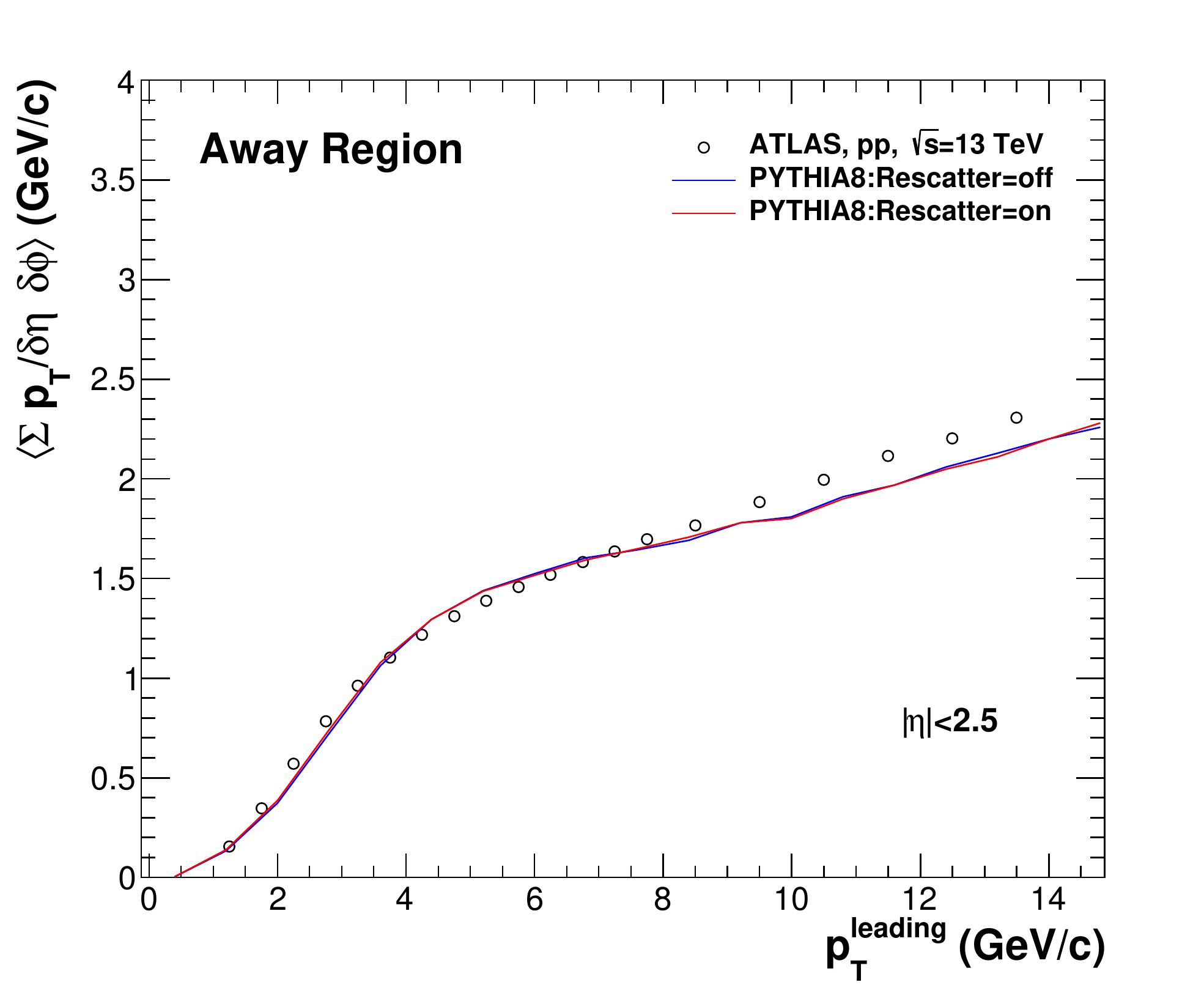}
\includegraphics[width=0.5\linewidth]{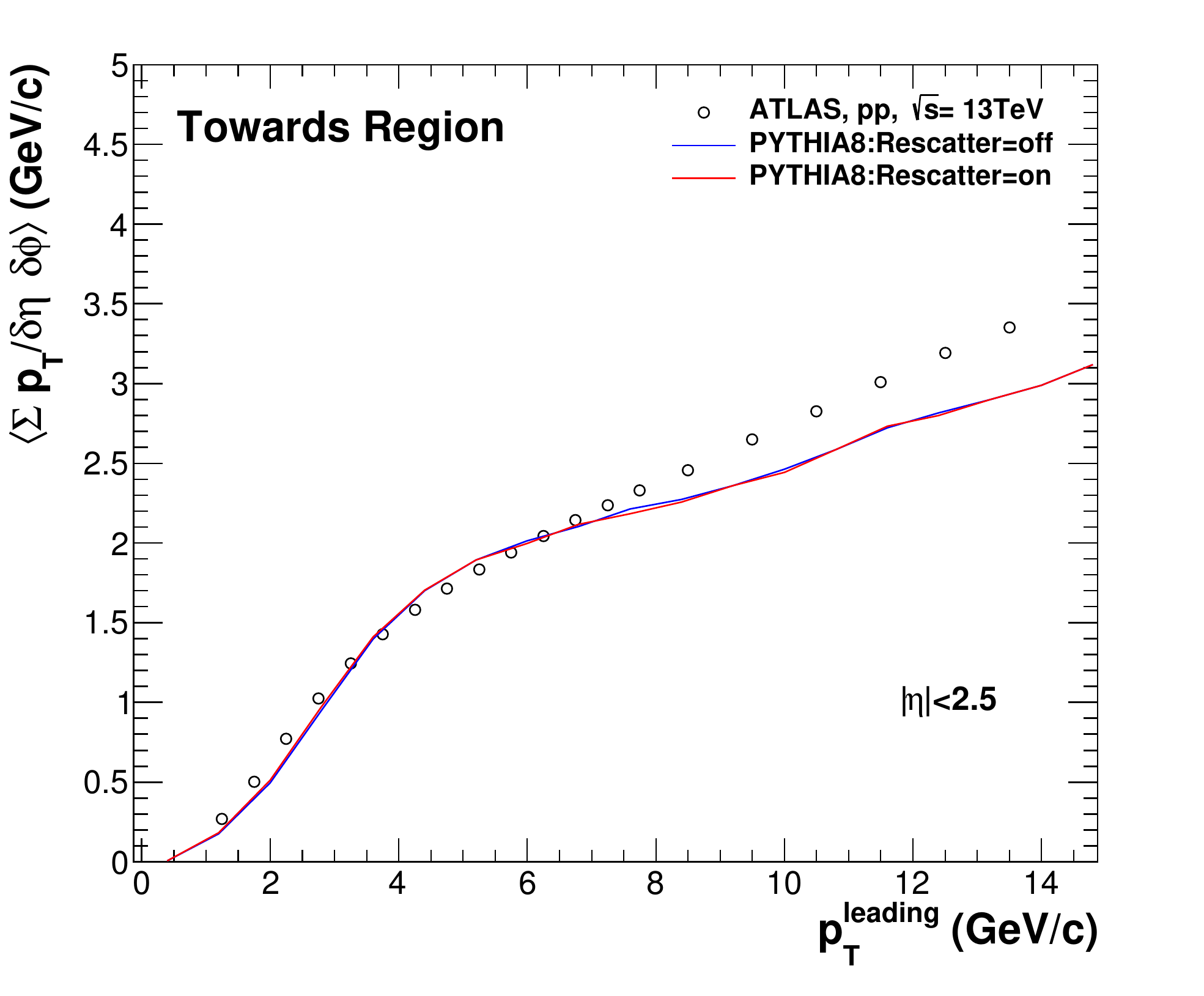}
\includegraphics[width=0.5\linewidth]{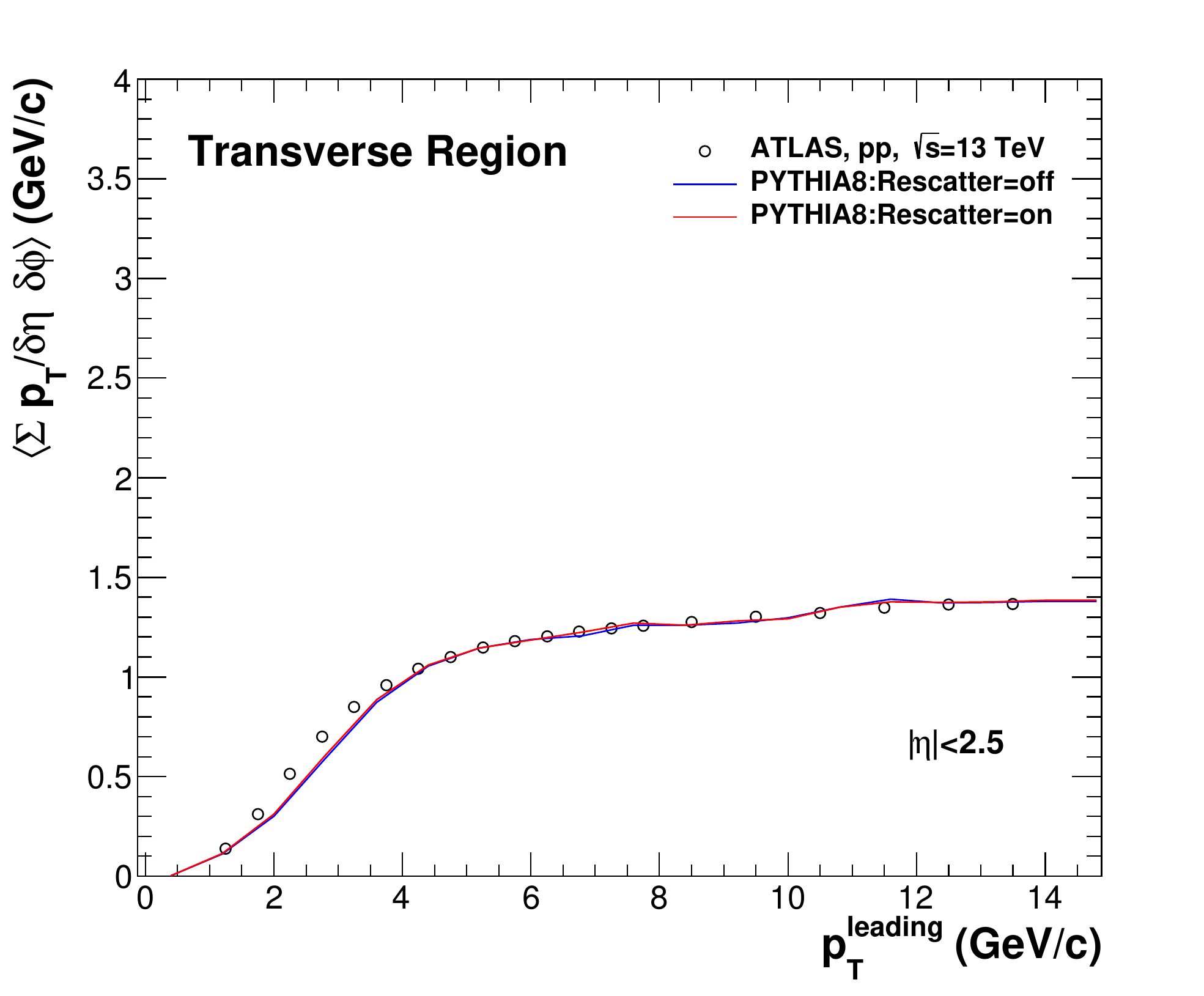}
\caption{ $\langle d^{2} \sum p_{T} /d\eta d\phi \rangle$as a function of  $p_{T}^{lead}$  for towards, away, and transverse  regions in  p$-$p collisions at $\sqrt{s}$ = 13 TeV regions with(and without) the effect of hadronic re-scattering. }
\label{fig3}
\end{figure}

Figure \ref{fig2} and  \ref{fig3} shows the variation of   $\langle d^{2}N /d\eta d\phi \rangle$  and  $\langle d^{2} \sum p_{T} /d\eta d\phi \rangle$ as a function of $p_{T}^{lead}$ for p$-$p collisions at  $\sqrt{s}$ = 13 TeV. The Pythia 8 predictions are compared with the ATLAS measurements \cite{atlas2}. Recently, the hadronic re-scattering has been introduced in Pythia 8 to study the effect of re-scattering of final state hadrons on final observables \cite{pythiarescatter}.  The figures also depict the predictions with (and without)  hadronic re-scattering. The general shape of the  evolution is similar for the three regions i.e. it exhibits a characteristic rapid rise for low values of $p_{T}^{lead}$  with an abrupt  change around 4.5 GeV/c value of $p_{T}^{lead}$.  In the transverse region,
the curve almost saturates showing a plateau like behavior while for the other two regions, it shows an increasing trend. This point of transition is mostly attributed to  reduction in impact
 parameter of p$-$p interactions and hence a change from soft to hard scattering regime. Therefore, one can see a continued activity from hard processes in towards and away region as the $p_{T}^{lead}$ increases
 while the transverse region is least affected.  This is consistent with the pedestal effect where the overlap between colliding protons is complete and any further growth seen is related to hard processes or contamination from the same rather than more MPI scattering.
As the toward region includes the leading charged particle, the multiplicity is slightly lower compared to away region as there is less energy available for additional particle production.  
The increase of the $\sum p_{T}$ densities in the toward and away regions indicates towards the partitioning of the total energy in each region in terms of charged particle production. 
The toward and away regions are the active regions compared to the transverse for higher values of $p_{T}^{lead}$.
The effect of hadronic re-scattering for the $\langle d^{2}N /d\eta d\phi \rangle$  is negligible at low $p_{T}^{lead}$ while for higher values it describes the data better in towards and away region. 
In the transverse region, the hadronic-rescattering over predicts the measured data. There is no effect of the hadronic  re-scattering for  $\langle d^{2} \sum p_{T} /d\eta d\phi \rangle$.\\

\begin{figure}
\centering
\includegraphics[width=0.7\linewidth]{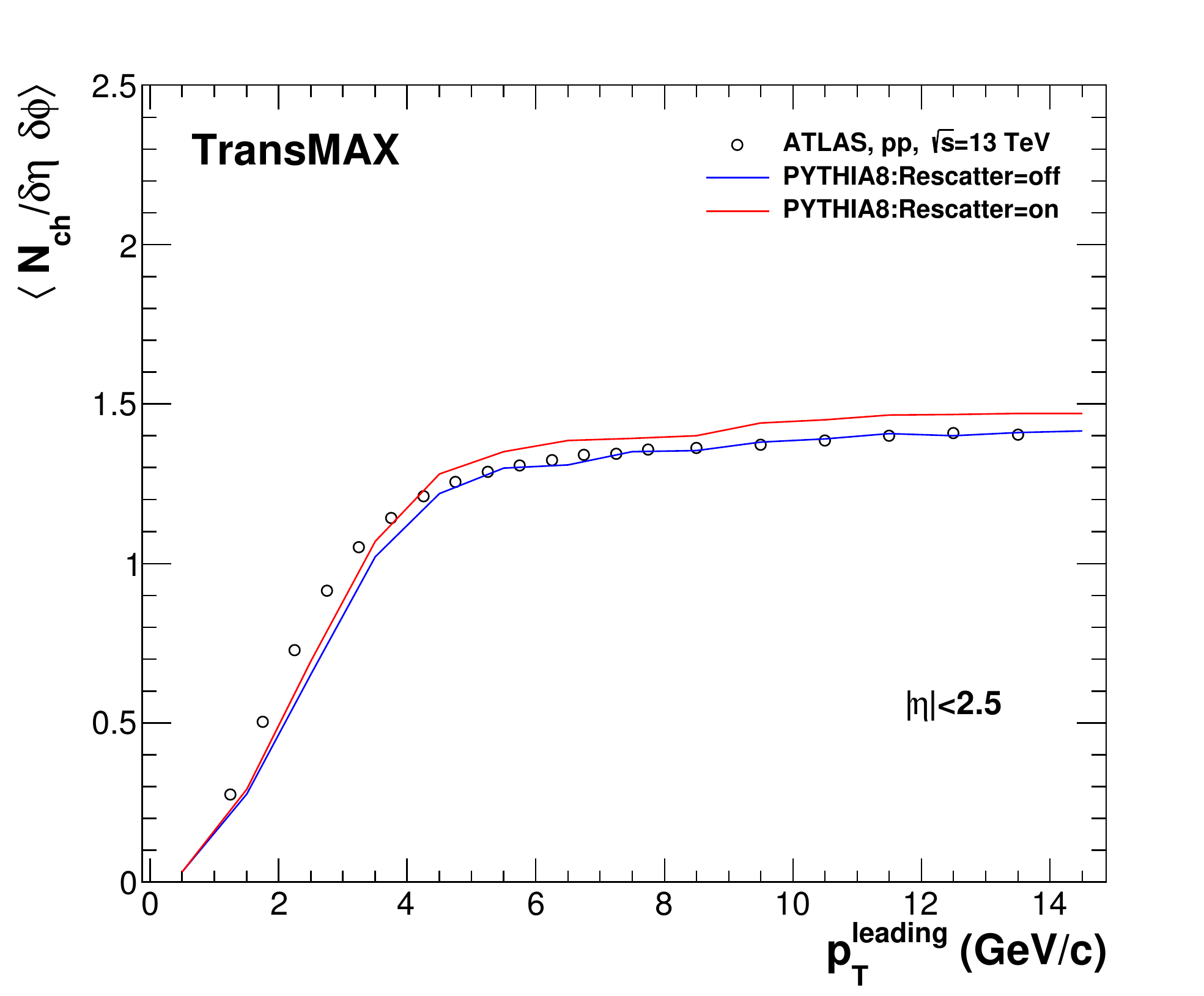}
\includegraphics[width=0.7\linewidth]{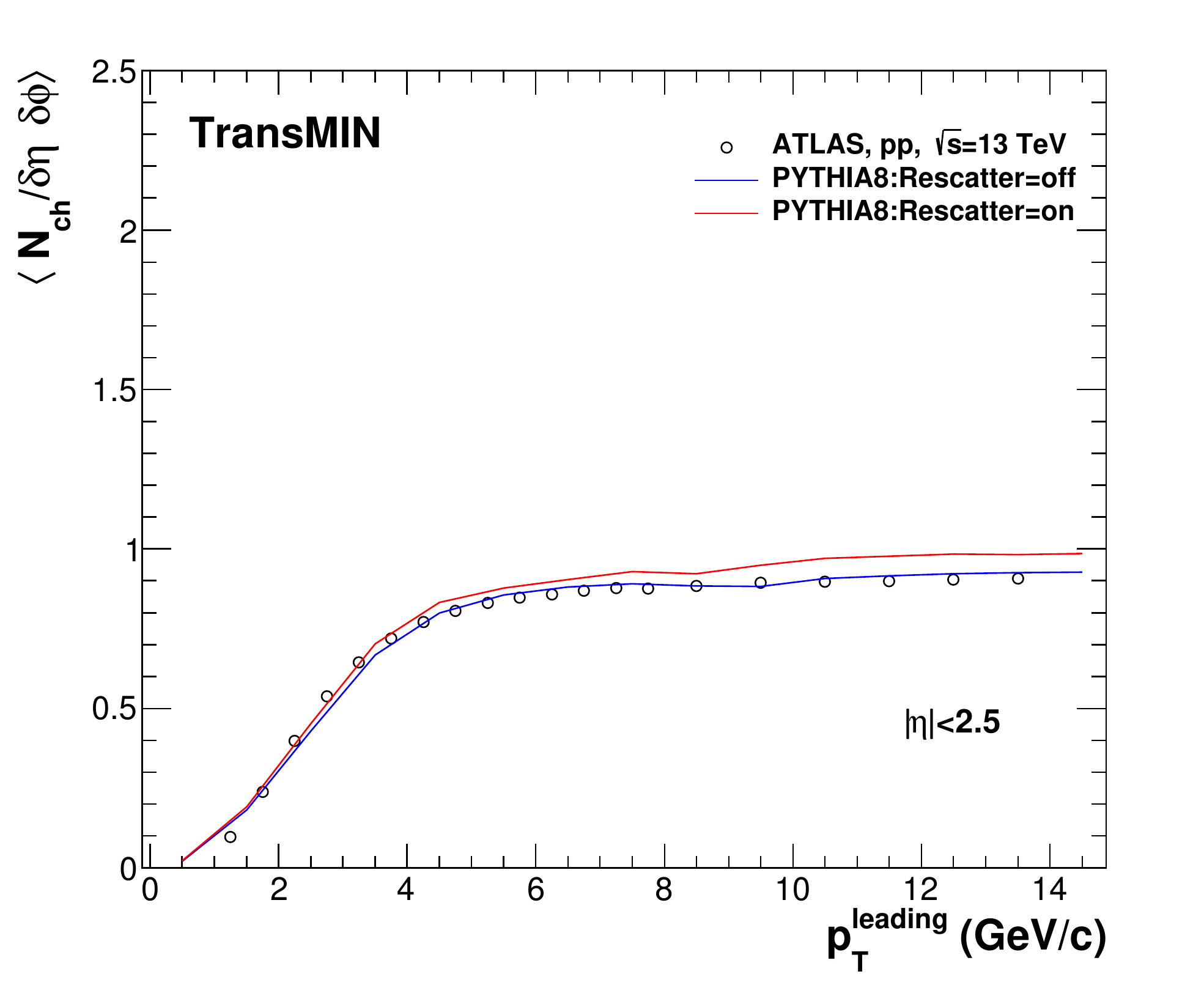}
\caption{$\langle d^{2}N /d\eta d\phi \rangle$ as a function of  $p_{T}^{lead}$  for TransMAX and TransMIN regions in  p$-$p collisions at $\sqrt{s}$ = 13 TeV regions with(and without) the effect of hadronic re-scattering. }
\label{fig4}
\end{figure}

\begin{figure}
\centering
\includegraphics[width=0.7\linewidth]{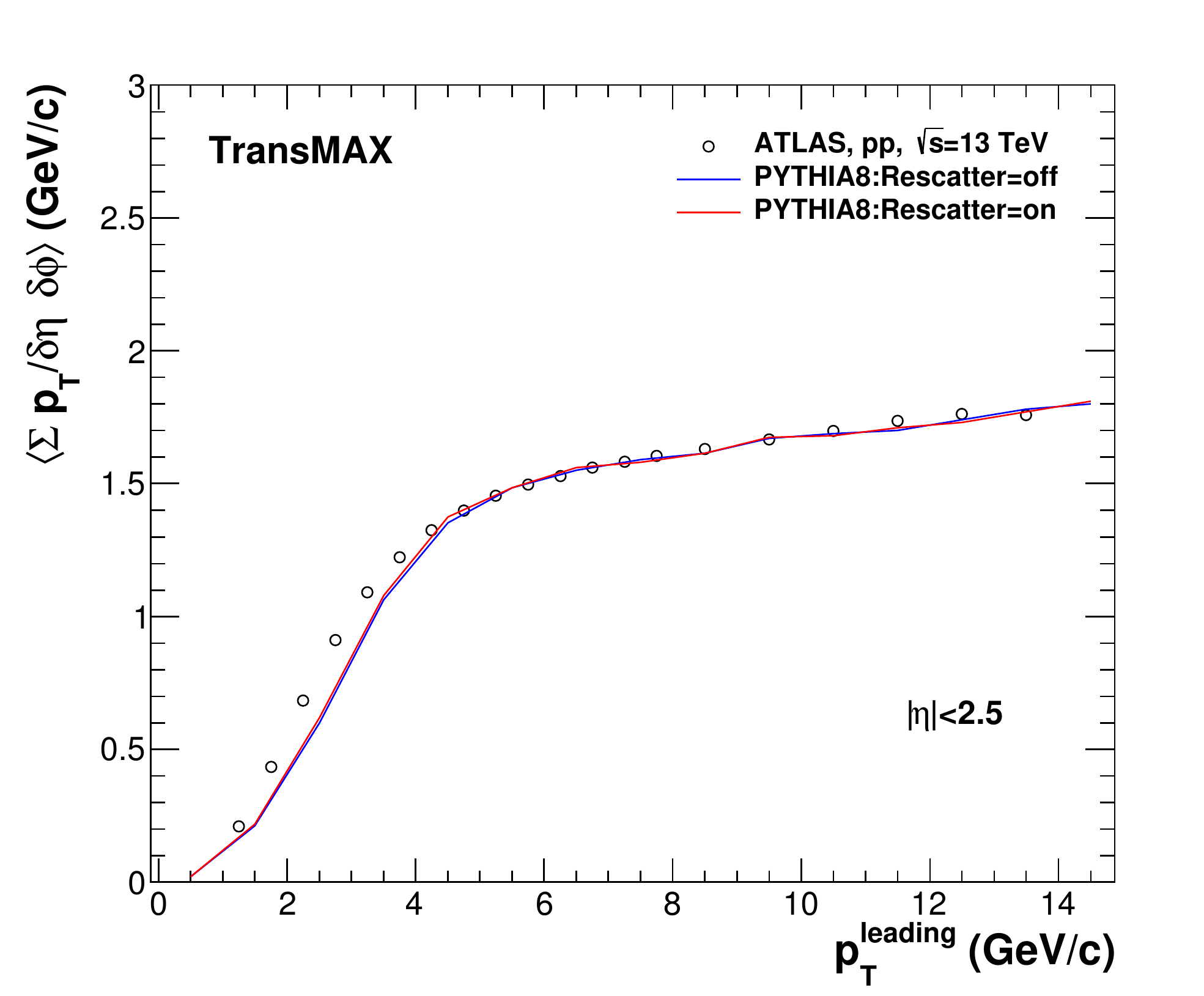}
\includegraphics[width=0.7\linewidth]{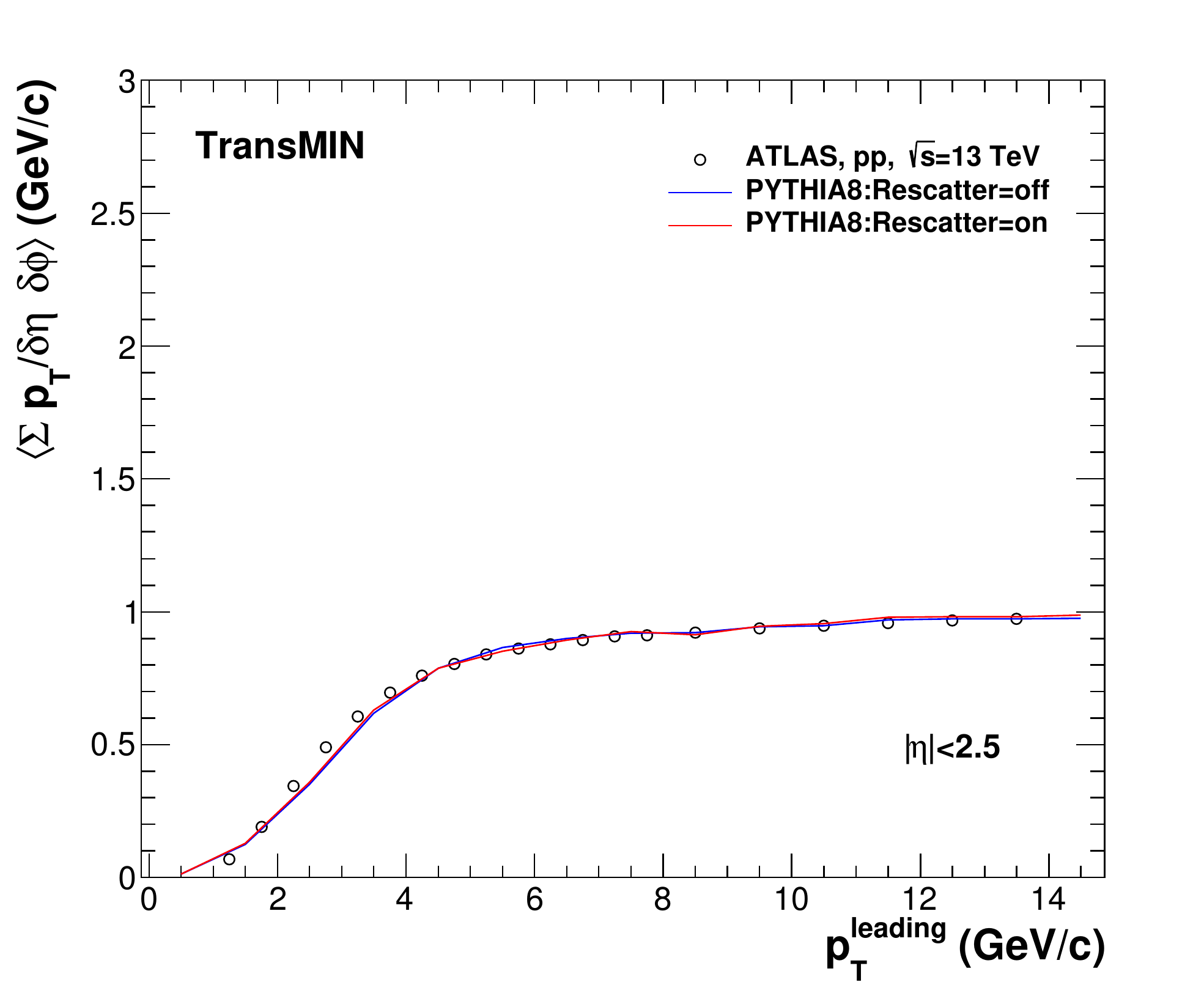}
\caption{ $\langle d^{2} \sum p_{T} /d\eta d\phi \rangle$as a function of  $p_{T}^{lead}$  for TransMAX and TransMIN regions in  p$-$p collisions at $\sqrt{s}$ = 13 TeV regions with(and without) the effect of hadronic re-scattering. }
\label{fig5}
\end{figure}

Figure  \ref{fig4}  and \ref{fig5} shows $\langle d^{2}N /d\eta d\phi \rangle$  and  $\langle d^{2} \sum p_{T} /d\eta d\phi \rangle$  as a function of $p_{T}^{lead}$  in TransMAX and TransMIN regions. The TransMIN region is dominated by pedestal effect while the TransMAX region receives contributions from MPIs as well as contaminations from hard-scattering processes.  One can observe an increase of  
charged particle densities  in TransMAX region with  $p_{T}^{lead}$  as it receives contributions from hard processes and the hadronic re-scattering slightly over predicts the measured data. Similarly, for mean $\sum p_{T}$ densities, one can observe a slightly increasing trend in activity with $p_{T}^{lead}$ for TransMAX region while it shows no activity  in the TransMIN region with no effect of hadronic re-scattering in either of the regions.

\begin{figure}
\centering
\includegraphics[width=0.5\linewidth]{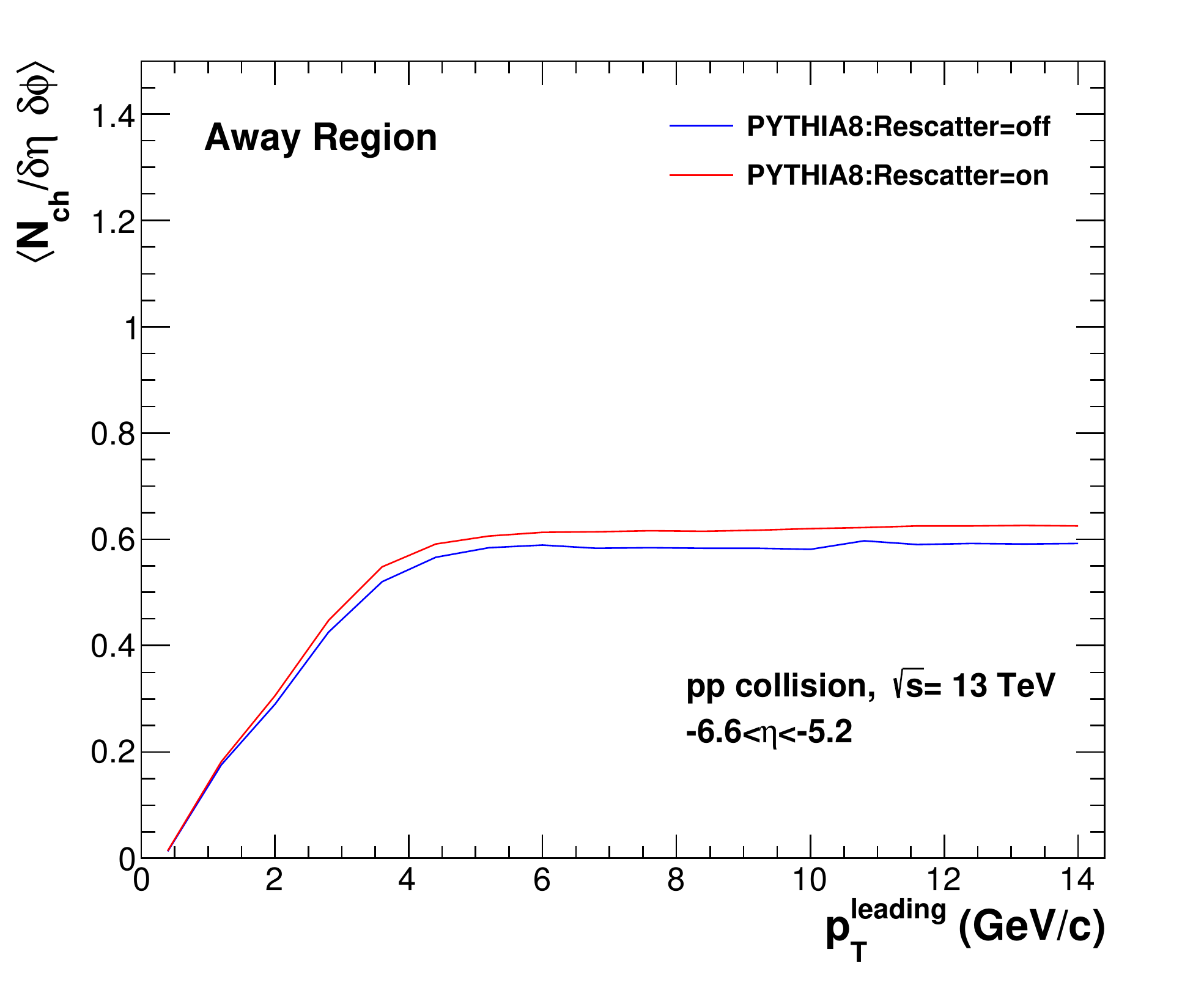}
\includegraphics[width=0.5\linewidth]{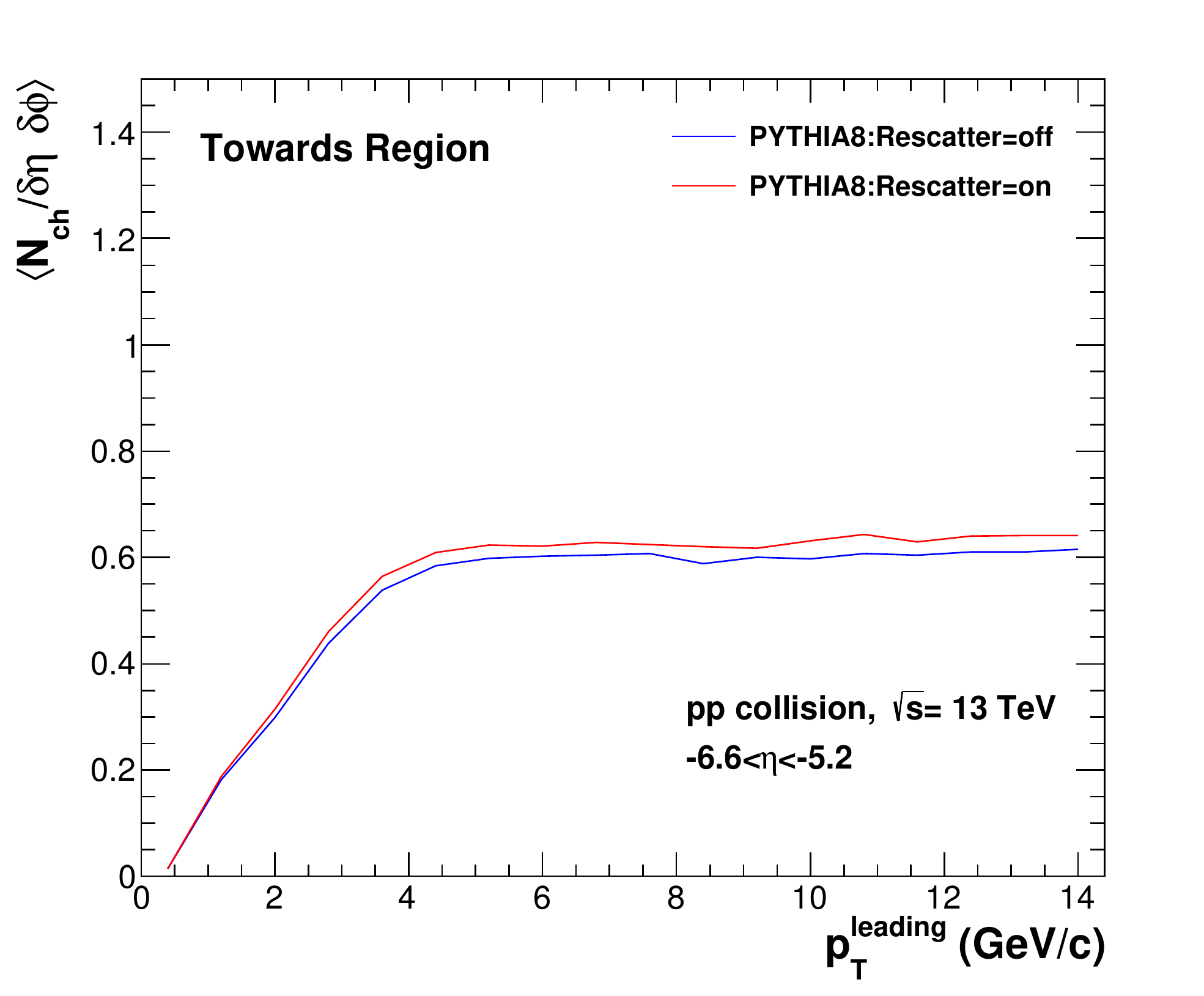}
\includegraphics[width=0.5\linewidth]{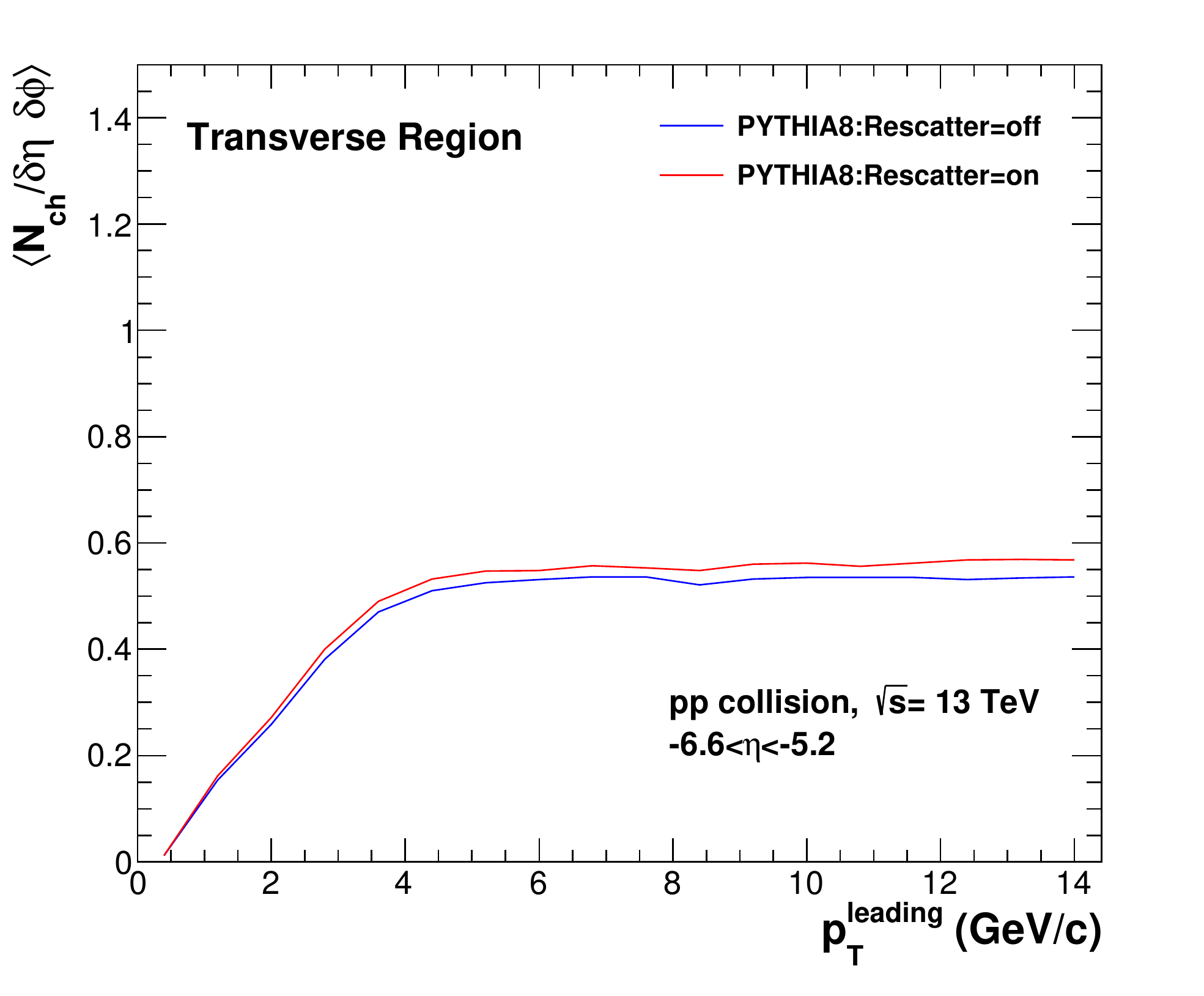}
\caption{$\langle d^{2}N /d\eta d\phi \rangle$ as a function of  $p_{T}^{lead}$  for towards, away, and transverse  regions in  p$-$p collisions at $\sqrt{s}$ = 13 TeV  for  -6.6 $< \eta <$ -5.2 with(and without) the effect of hadronic re-scattering. }
\label{fig6}
\end{figure}

\begin{figure}
\centering
\includegraphics[width=0.5\linewidth]{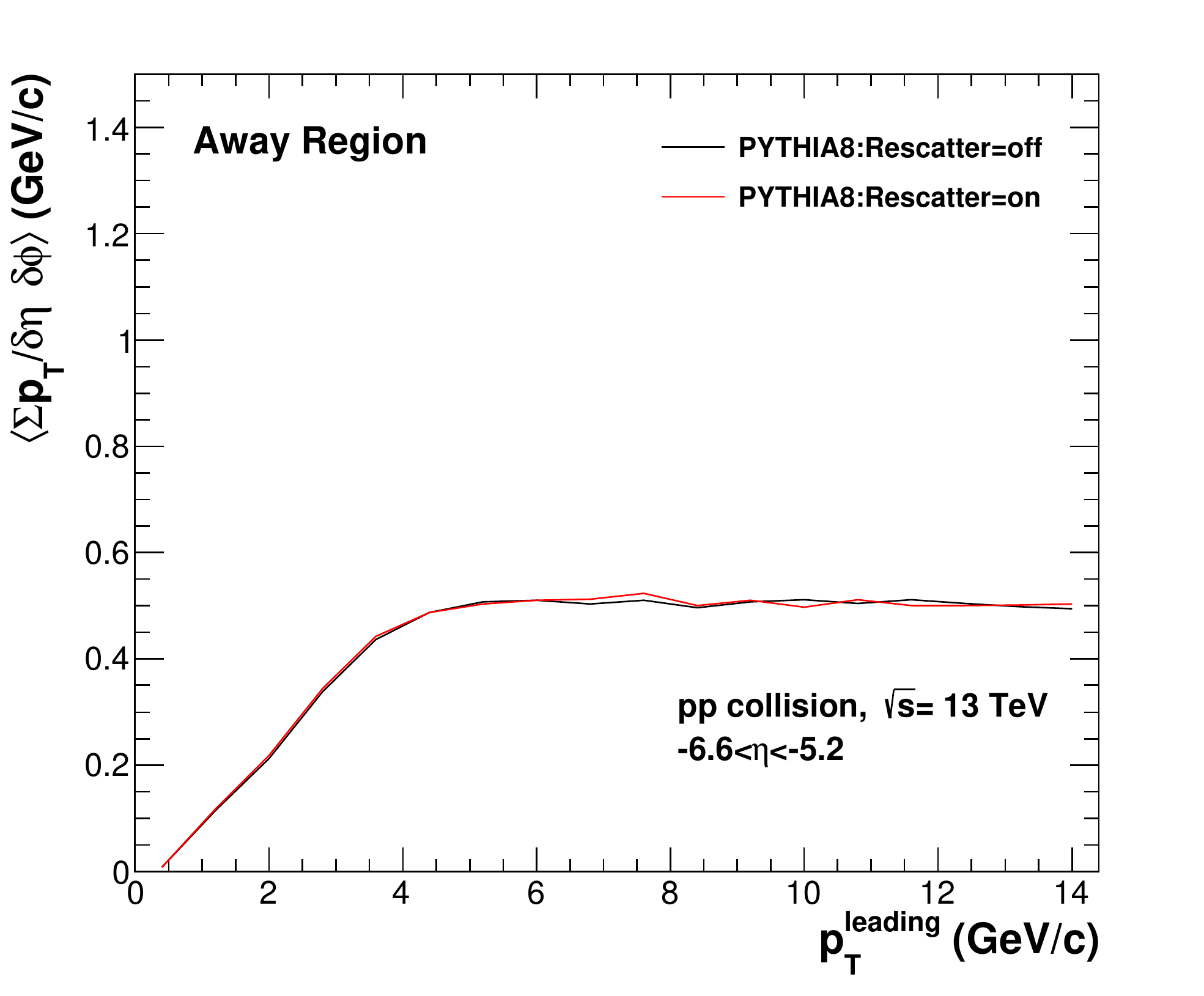}
\includegraphics[width=0.5\linewidth]{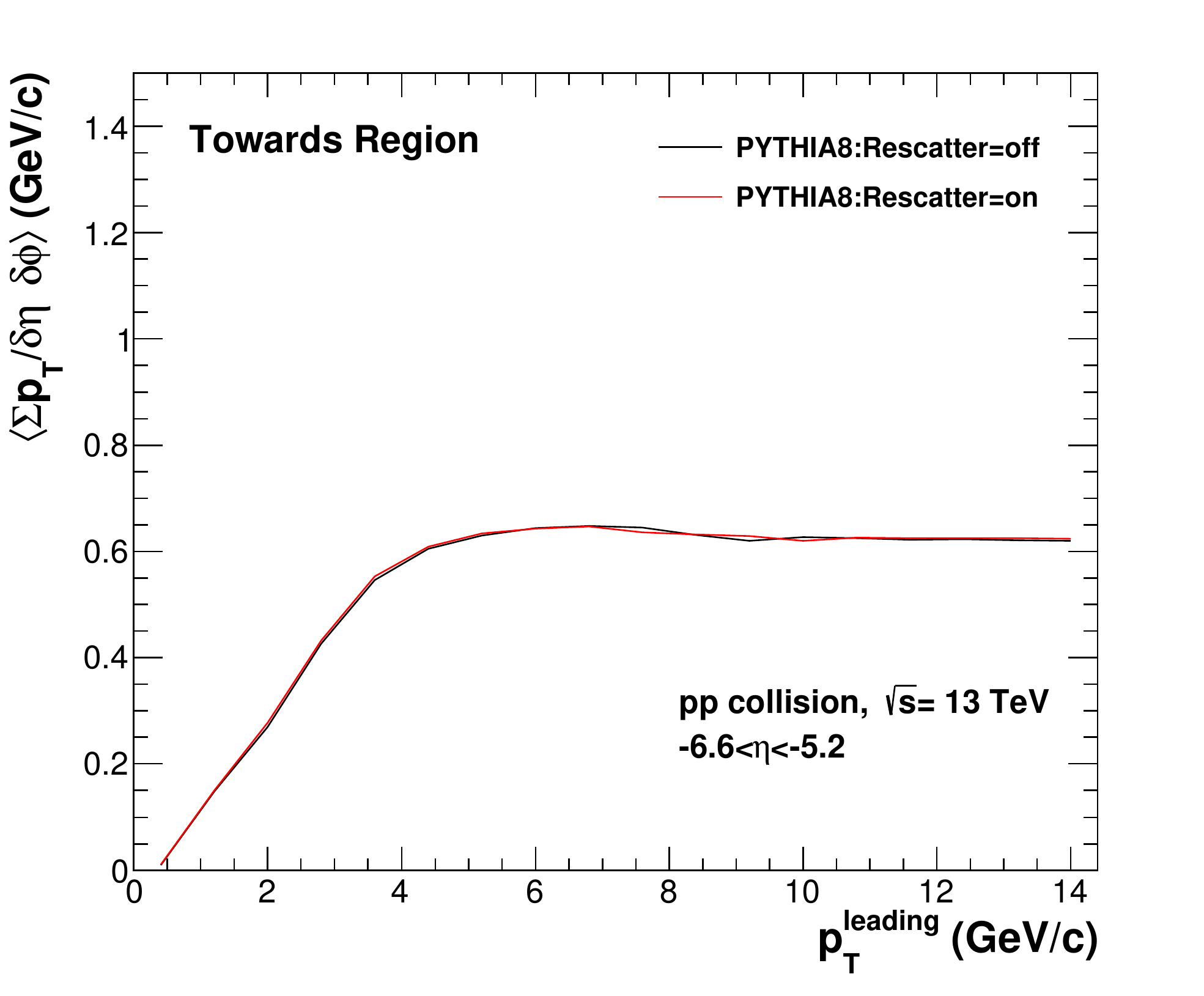}
\includegraphics[width=0.5\linewidth]{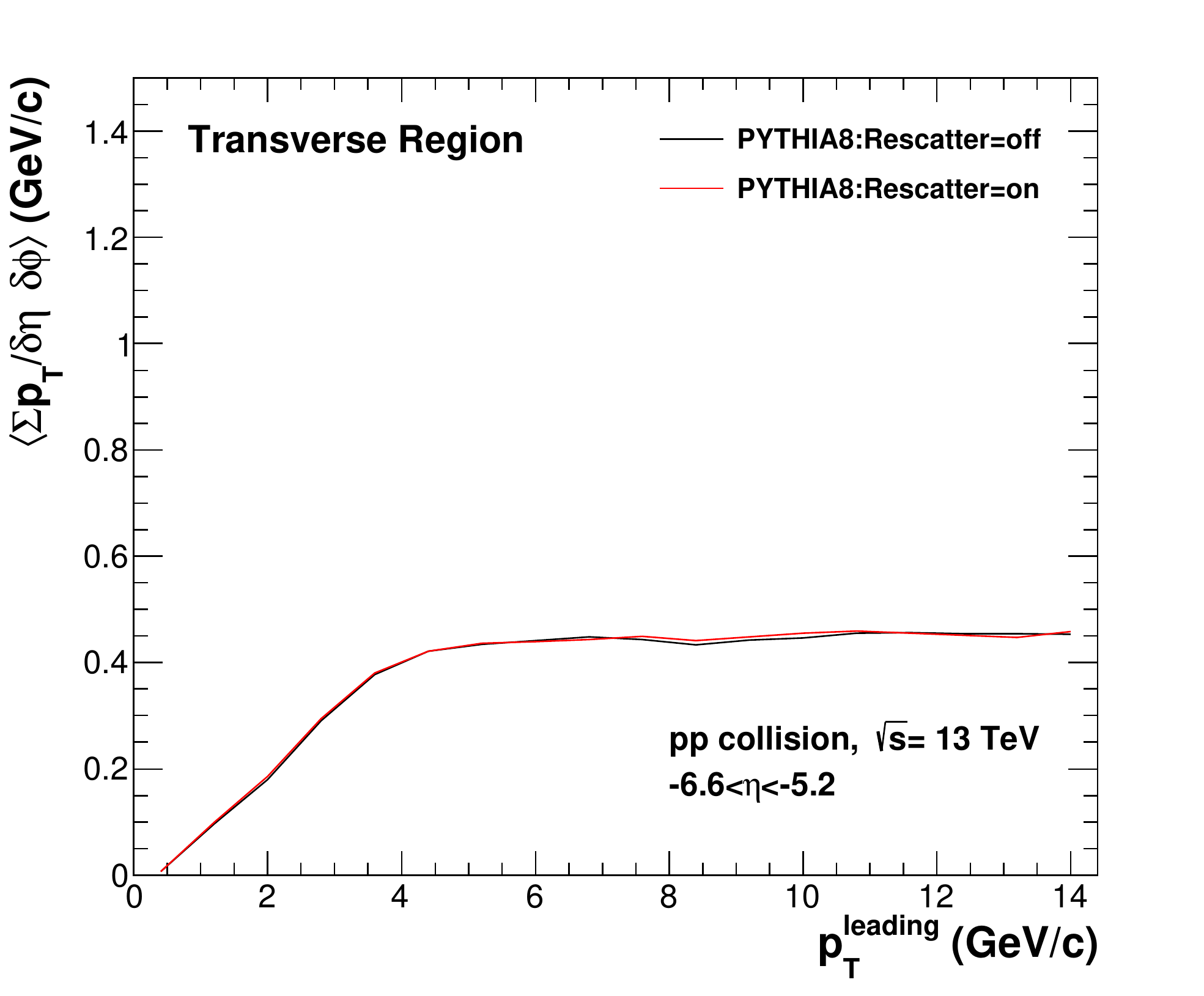}
\caption{ $\langle d^{2} \sum p_{T} /d\eta d\phi \rangle$as a function of  $p_{T}^{lead}$  for towards, away, and transverse  regions in  p$-$p collisions at $\sqrt{s}$ = 13 TeV for  -6.6 $< \eta <$ -5.2 with(and without) the effect of hadronic re-scattering. }
\label{fig7}
\end{figure}

\begin{figure}
\centering
\includegraphics[width=0.7\linewidth]{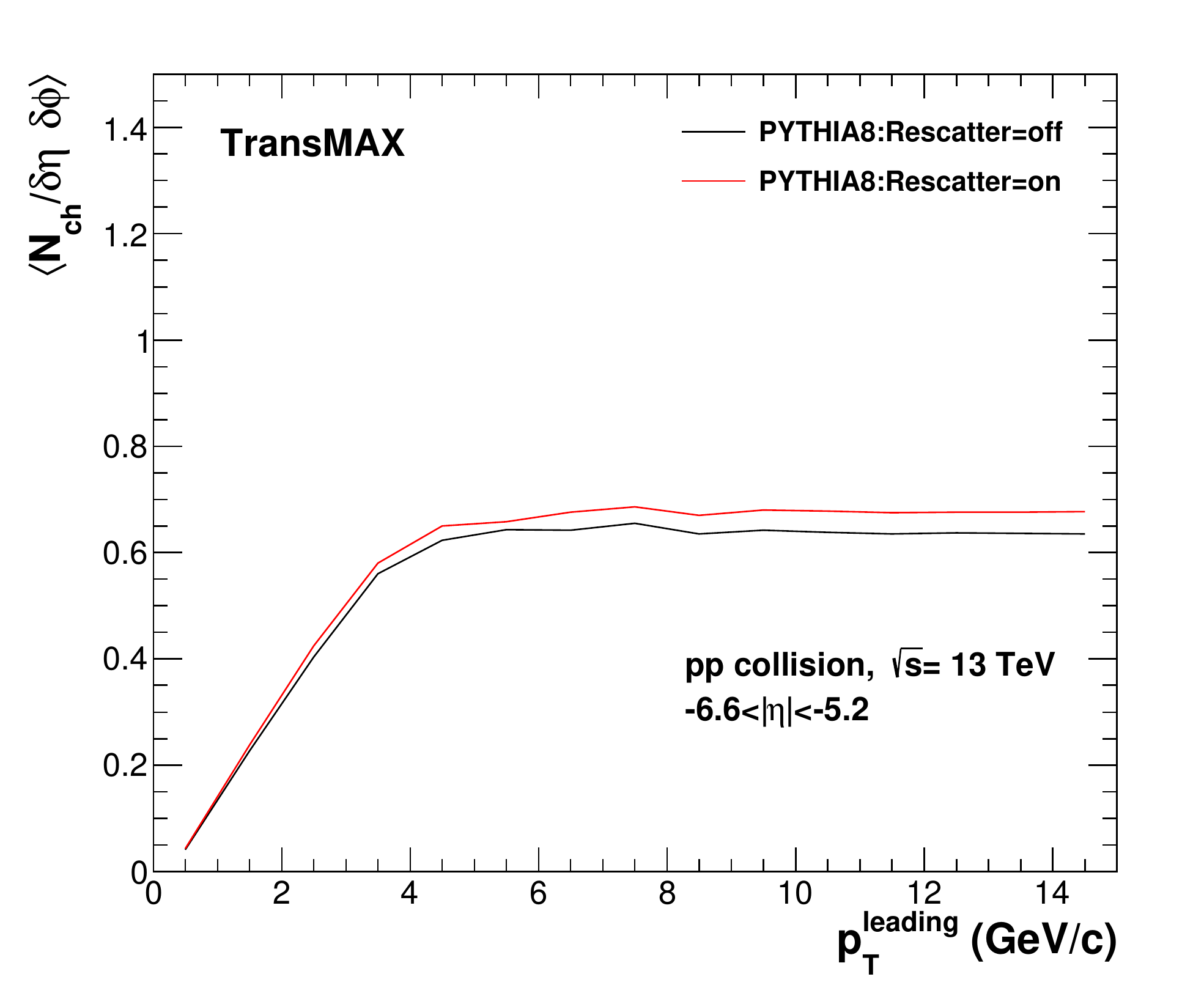}
\includegraphics[width=0.7\linewidth]{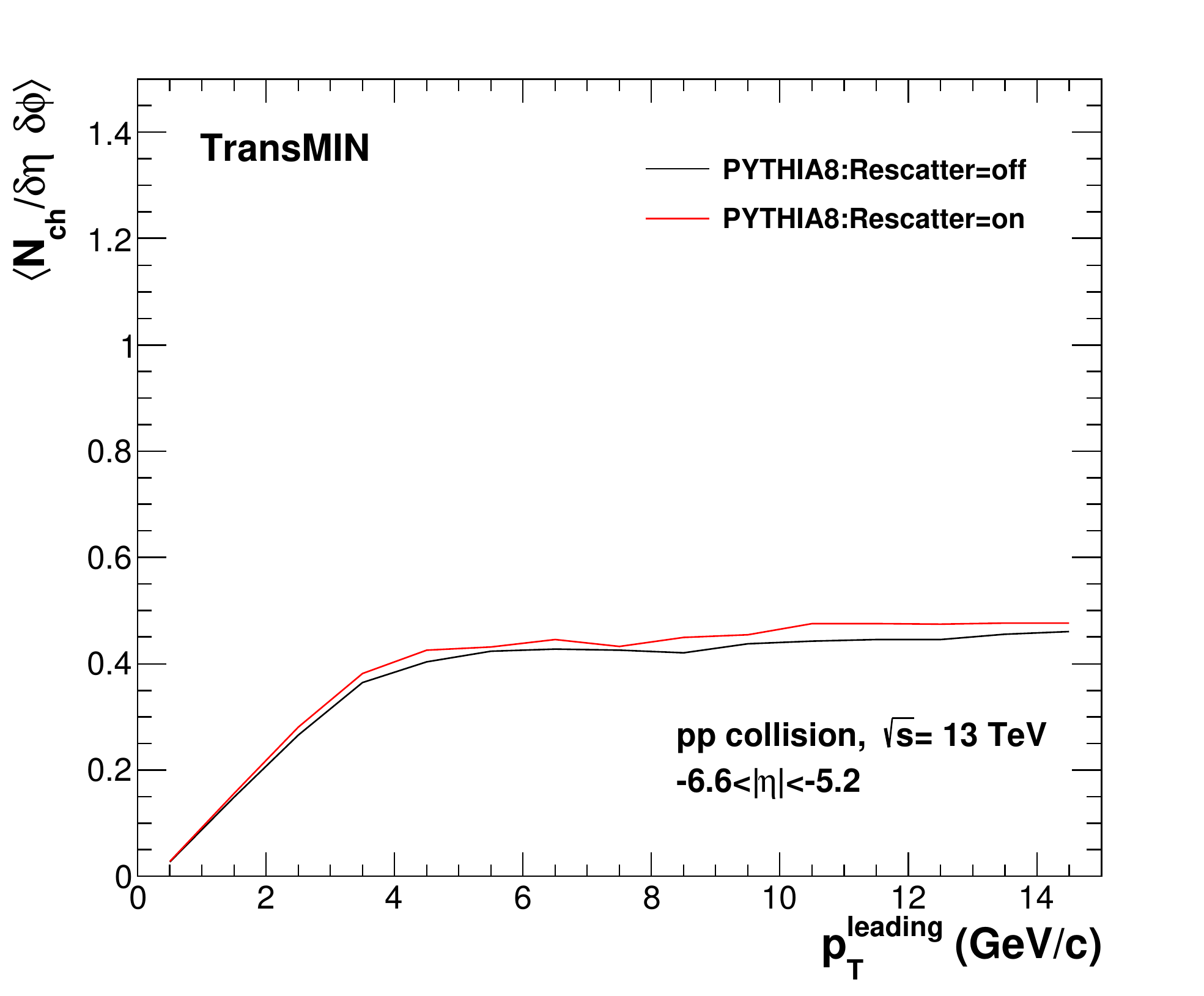}
\caption{$\langle d^{2}N /d\eta d\phi \rangle$ as a function of  $p_{T}^{lead}$  for TransMAX and TransMIN regions in  p$-$p collisions at $\sqrt{s}$ = 13 TeV for  -6.6 $< \eta <$ -5.2   with(and without) the effect of hadronic re-scattering. }
\label{fig8}
\end{figure}

\begin{figure}
\centering
\includegraphics[width=0.7\linewidth]{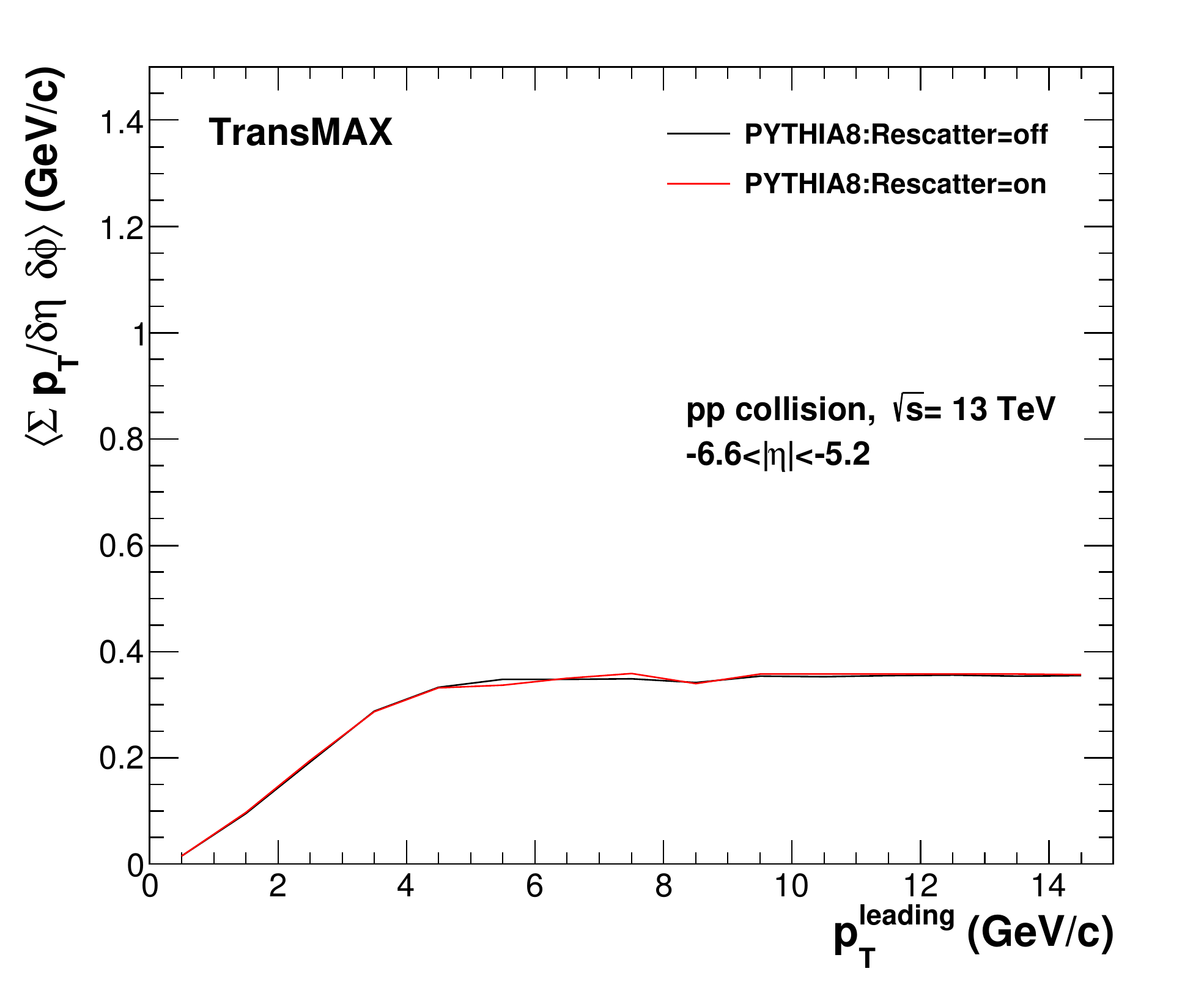}
\includegraphics[width=0.7\linewidth]{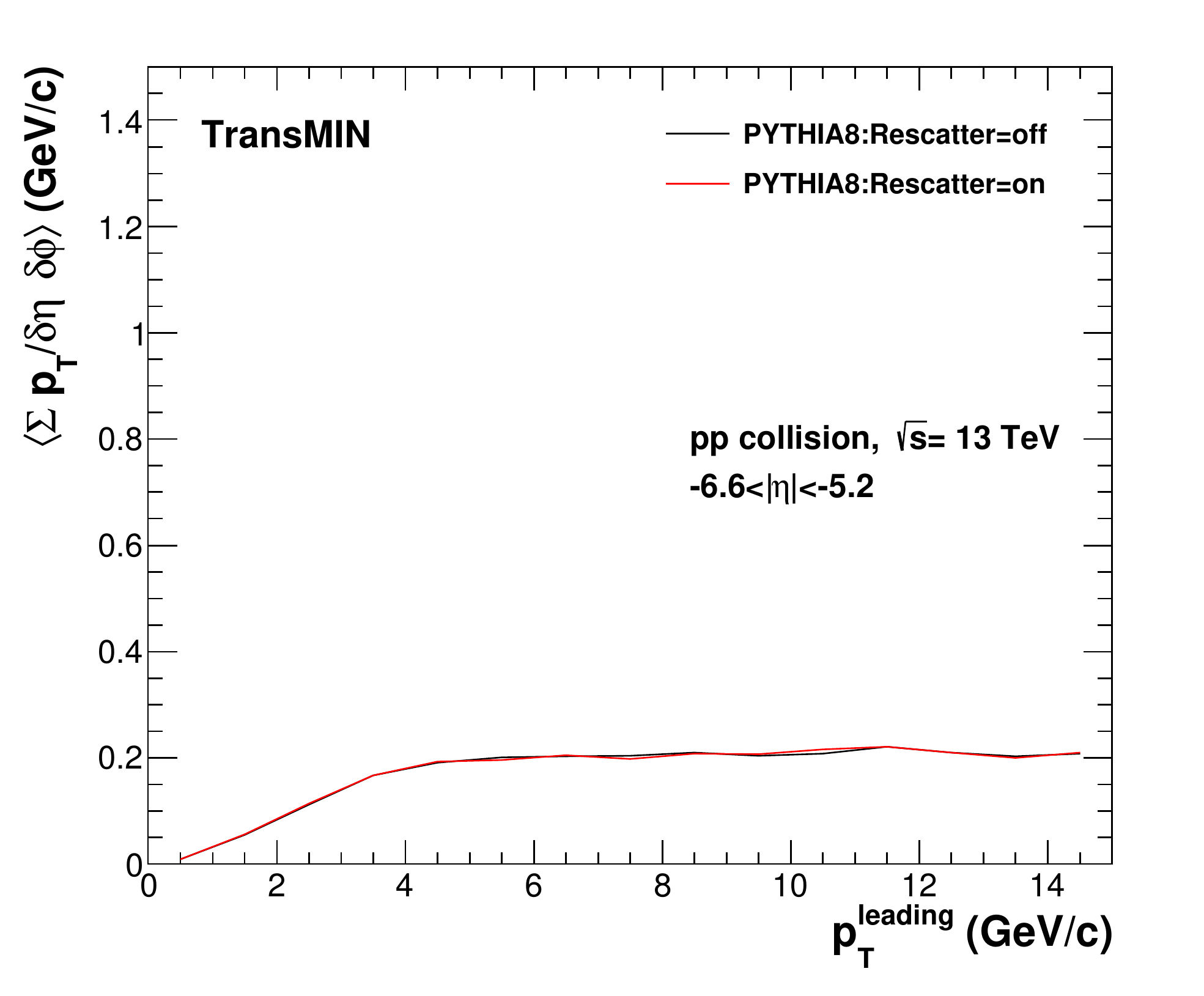}
\caption{ $\langle d^{2} \sum p_{T} /d\eta d\phi \rangle$ as a function of  $p_{T}^{lead}$  for TransMAX and TransMIN regions in  p$-$p collisions at $\sqrt{s}$ = 13 TeV for  -6.6 $< \eta <$ -5.2 with(and without) the effect of hadronic re-scattering. }
\label{fig9}
\end{figure}

The hadronic activity has been also studied  in forward regions (higher $\eta$ regions) where the contaminations from hard scattering  processes are expected to be minimum and all the three regions are therefore  
sensitive to underlying events. Figure  \ref{fig6}  and  \ref{fig7} shows  $\langle d^{2}N /d\eta d\phi \rangle$  and  $\langle d^{2} \sum p_{T} /d\eta d\phi \rangle$ as a function of  $p_{T}^{lead}$  in the forward region with  $-6.6  < \eta < -5.2$ . It can be observed that at low $p_{T}^{lead}$, there is a sharp rise in both $\langle d^{2}N /d\eta d\phi \rangle$  and  $\langle d^{2} \sum p_{T} /d\eta d\phi \rangle$ with $p_{T}^{lead}$  in towards, away, and transverse regions due to increased multi-partonic interactions but $\sim$  4.0 GeV/c onwards, both the observables saturates in all the three regions. This interesting observation of the plateau in towards and away regions is in contrast with the observation at central region. This indicates the near absence of contributions from hard -scattering processes in forward regions\cite{chatrchyan2013study}. The effect of hadronic re-scattering is similar, as discussed earlier, in the towards, away, and transverse regions in the forward pseudo-rapidity region as observed in central regions. 
Figure \ref{fig8} and \ref{fig9} shows the same in TransMAX and TransMIN regions at $\sqrt{s}$ = 13 TeV and the trend is similar to that observed in central regions.
 
\begin{figure}
\includegraphics[width=0.5\linewidth]{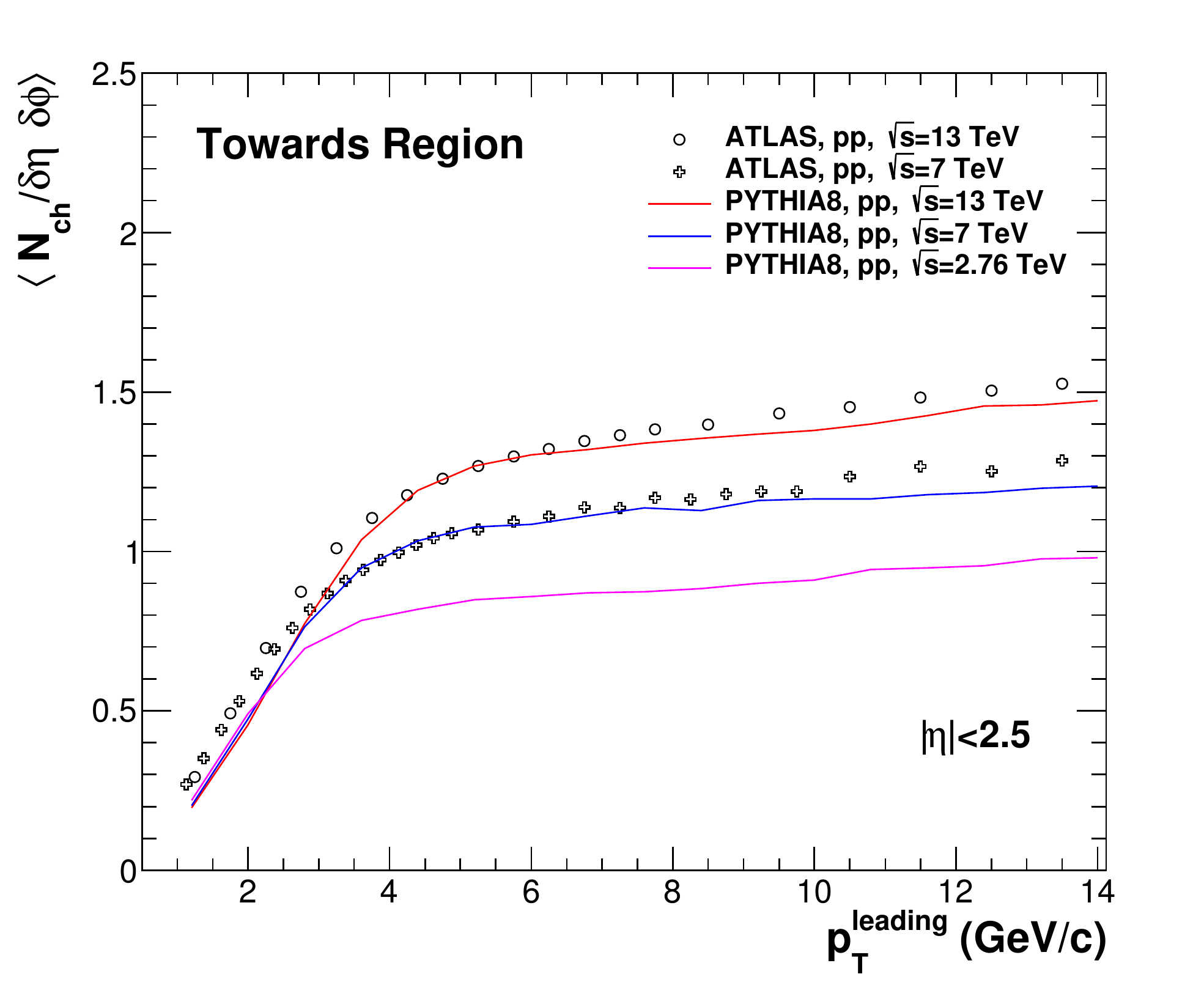}
\includegraphics[width=0.5\linewidth]{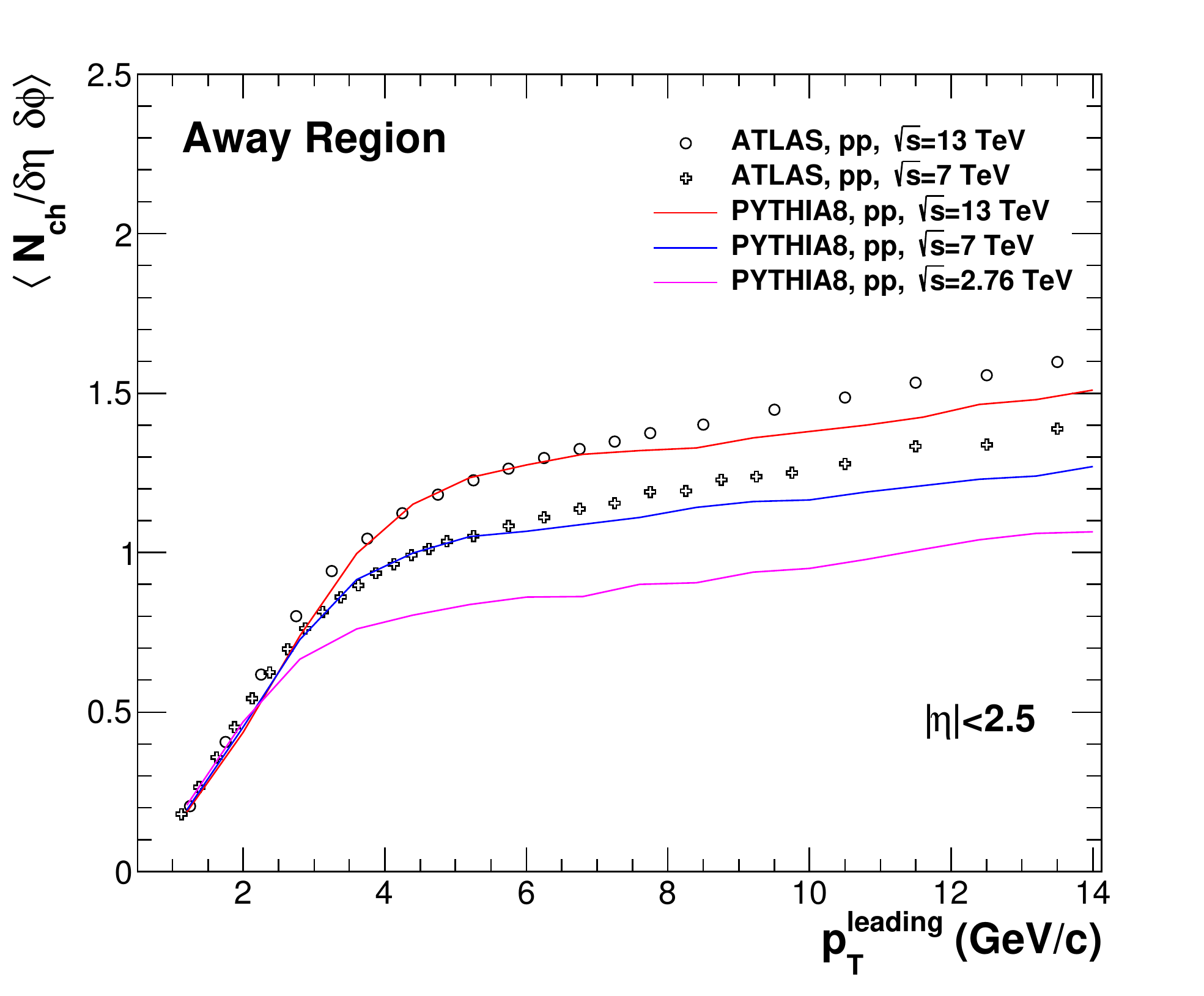}
\includegraphics[width=0.5\linewidth]{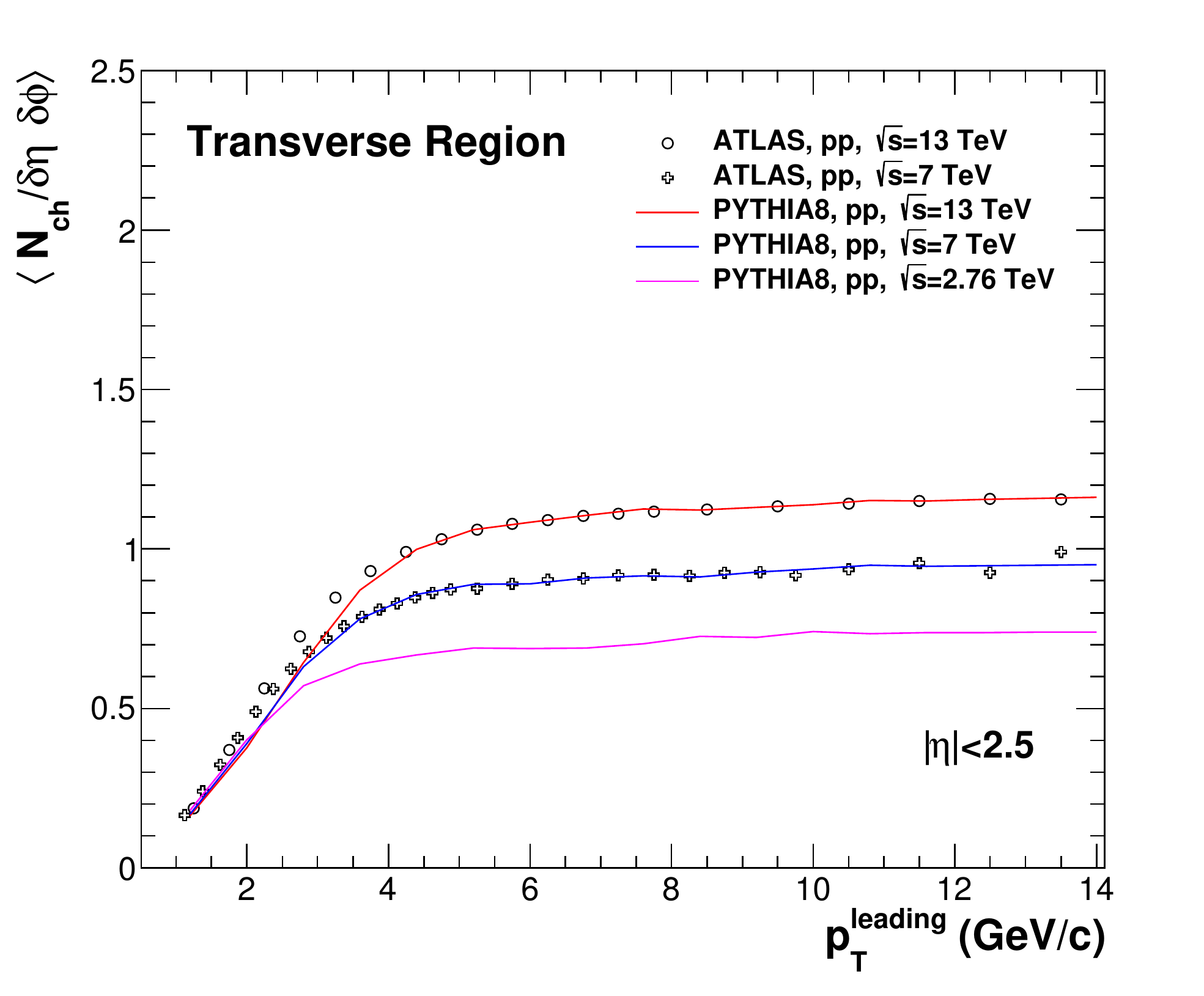}
\caption{ $\langle d^{2}N /d\eta d\phi \rangle$  as a function of $p_{T}^{lead}$ for towards, away, and transverse regions for  p$-$p collisions at $\sqrt{s}$ = 2.76, 7 and 13 TeV}
\label{fig10}
\end{figure}

\begin{figure}
\includegraphics[width=0.5\linewidth]{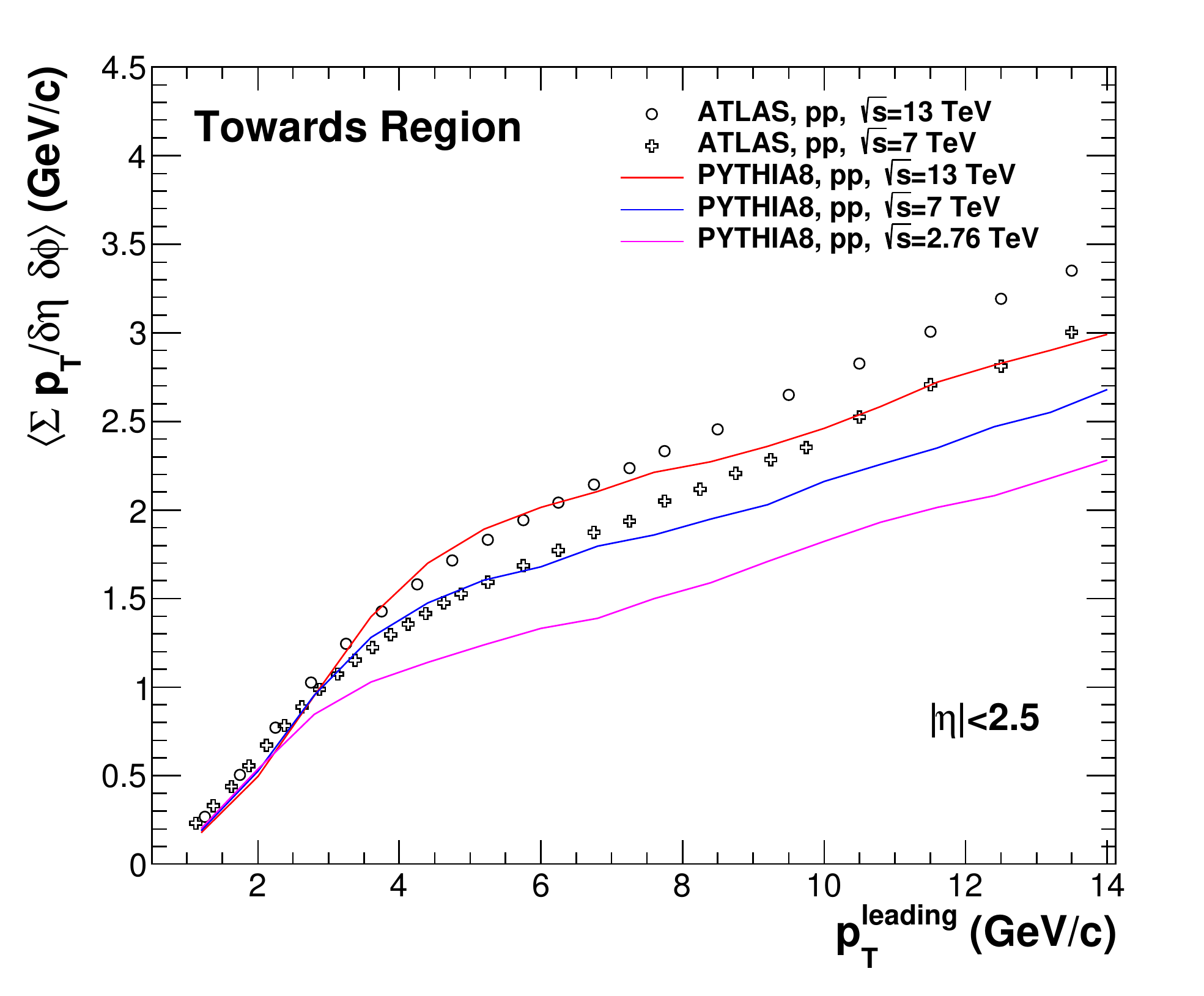}
\includegraphics[width=0.5\linewidth]{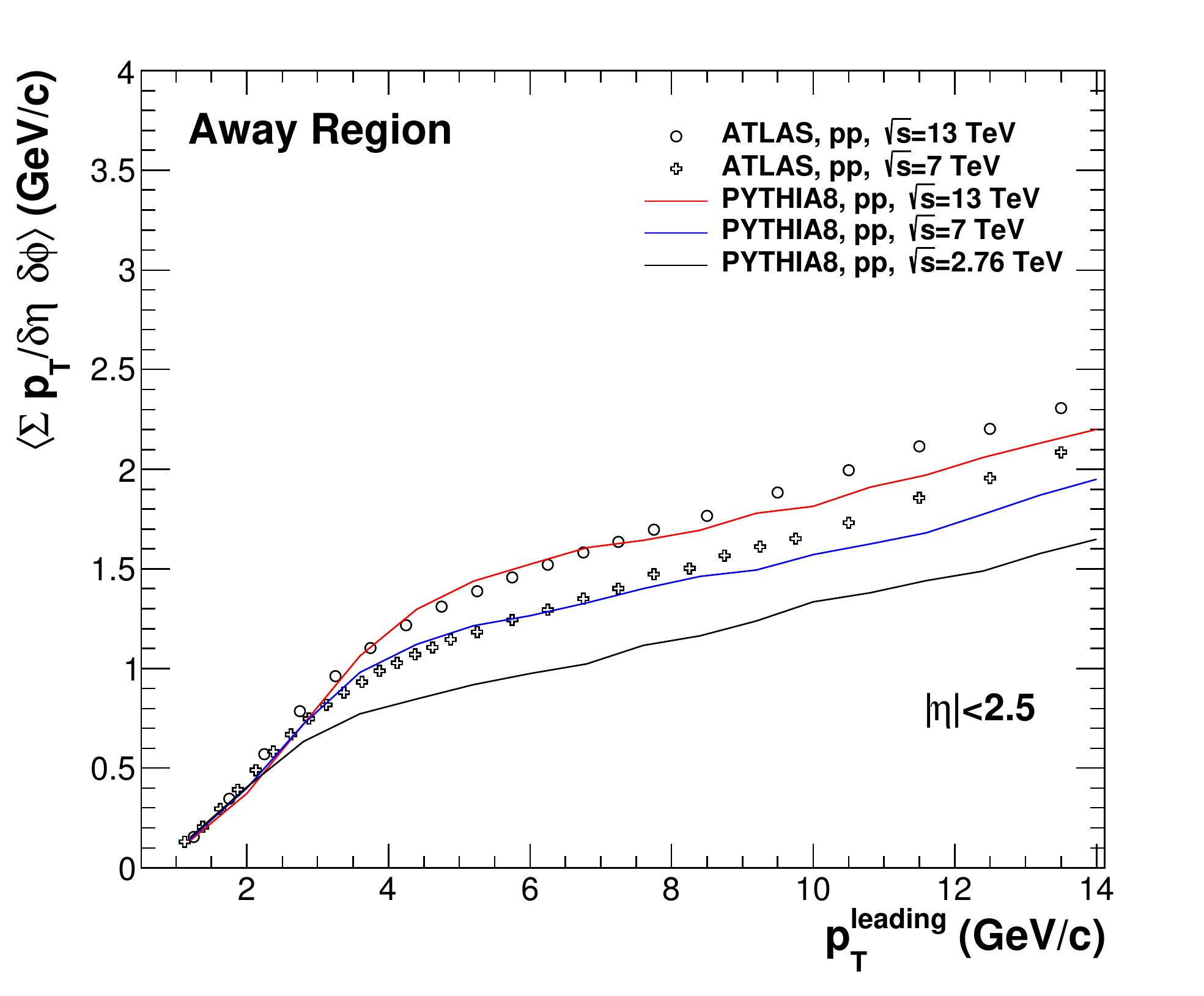}
\includegraphics[width=0.5\linewidth]{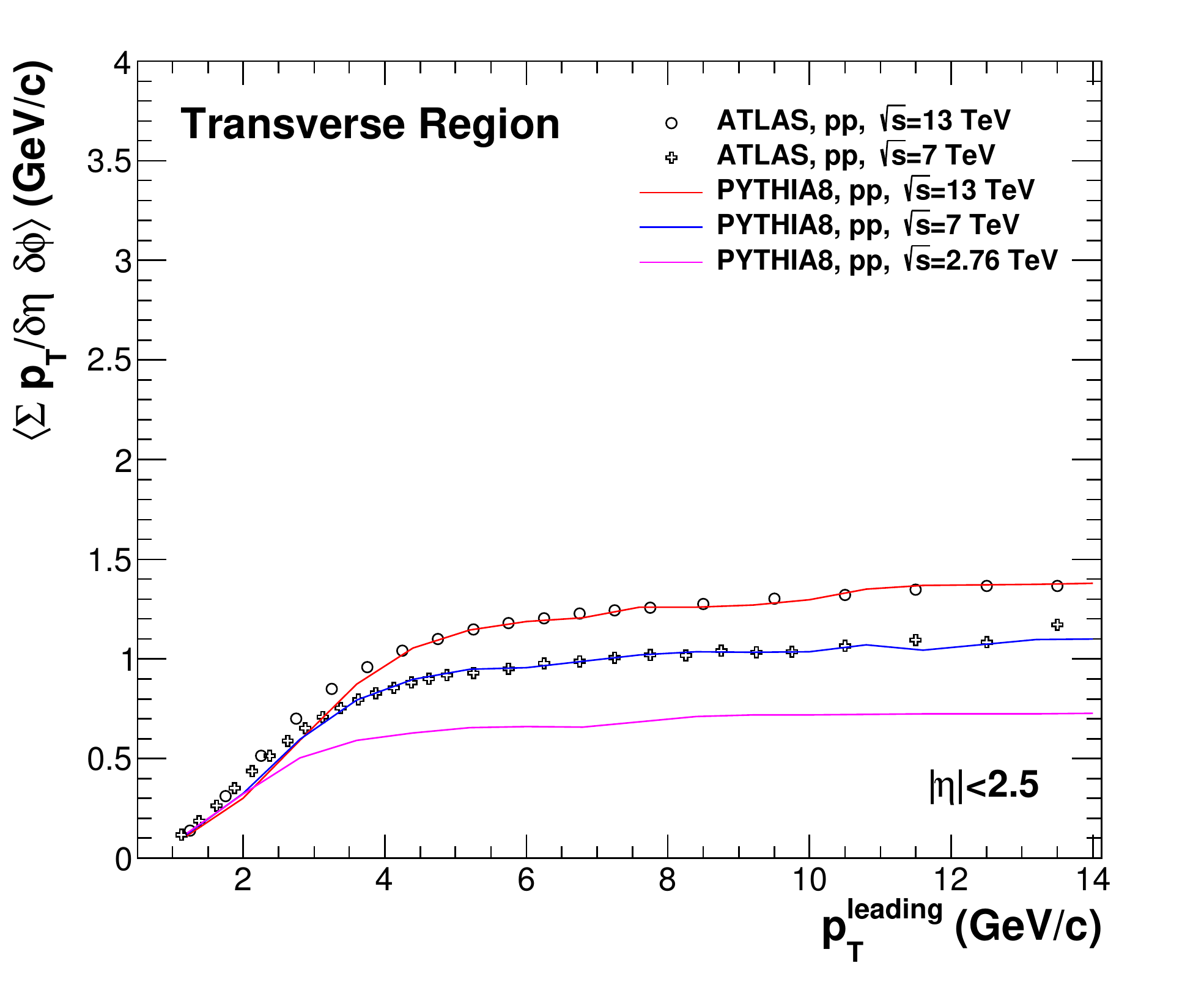}
\caption{ $\langle d^{2} \sum p_{T} /d\eta d\phi \rangle$  as a function  $p_{T}^{lead}$  for towards, away, and transverse regions for for  p$-$p collisions at $\sqrt{s}$ = 2.76, 7 and 13 TeV.}
\label{fig11}
\end{figure}

The energy dependence of the underlying events in p$-$p collisions have been studied at centre of mass-energy, $\sqrt{s}$ = 2.76, 7 and 13 TeV and have been shown in Figure \ref{fig10} and \ref{fig11}. One can observe that there is a  strong rise in mean charged  particle multiplicity  and mean $\sum p_{T}$ densities  with increasing $\sqrt{s}$ due to increased partonic event activities. It is also observed that  on going from 2.76  TeV to 13 TeV, the mean charge multiplicity and mean scalar $p_T$ sum is almost doubled, which implies that the multi-partonic interactions (MPI) activity grows more with centre of mass-energy\cite{cdfstudy,ortiz2017universality}. It was also shown in reference \cite{ortiz2017universality} that the dependence on beam energy  becomes vanishingly small as the charged particle densities are scaled by the relative rise in multiplicity.

\begin{figure}
\includegraphics[width=0.5\linewidth]{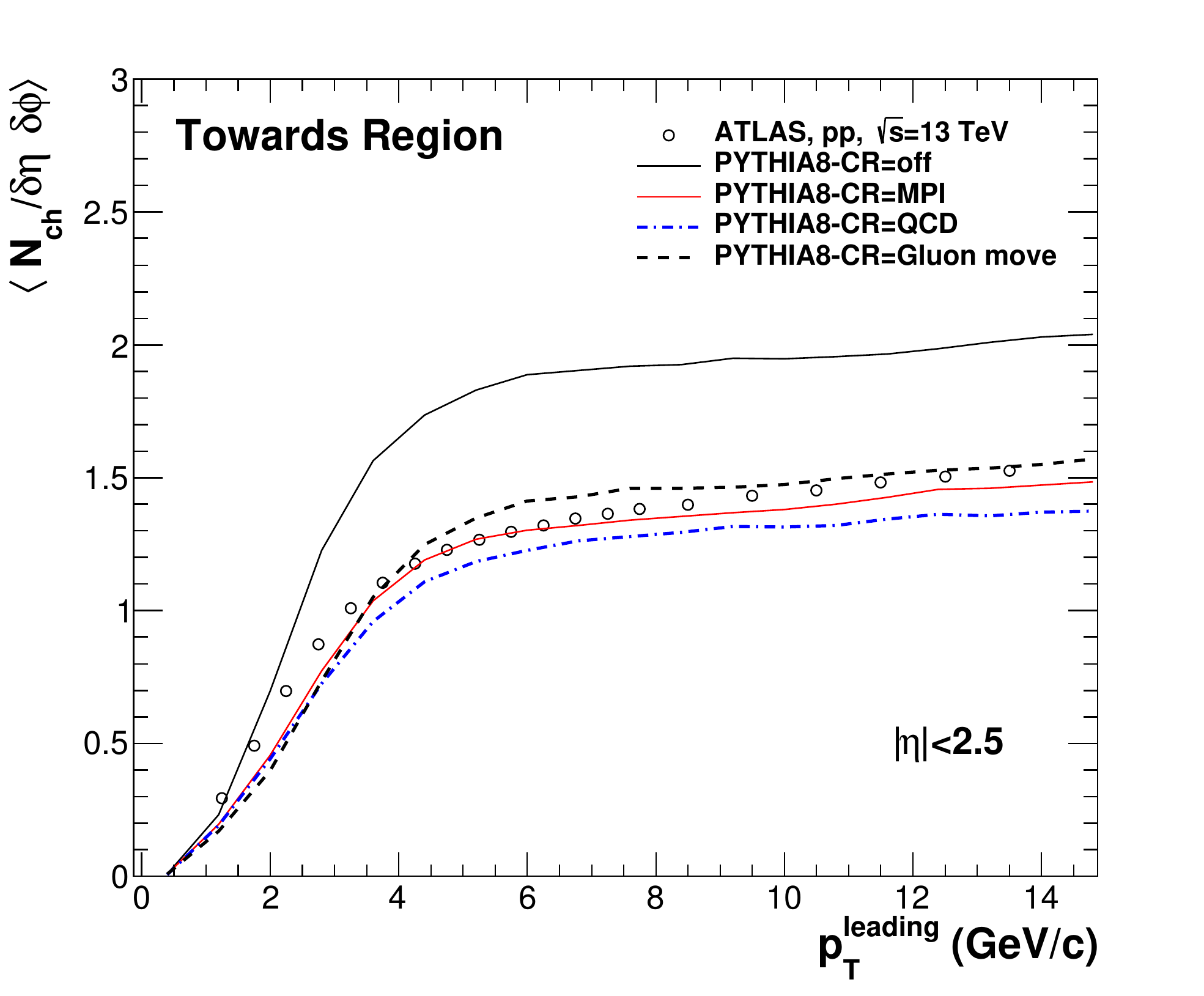}
\includegraphics[width=0.5\linewidth]{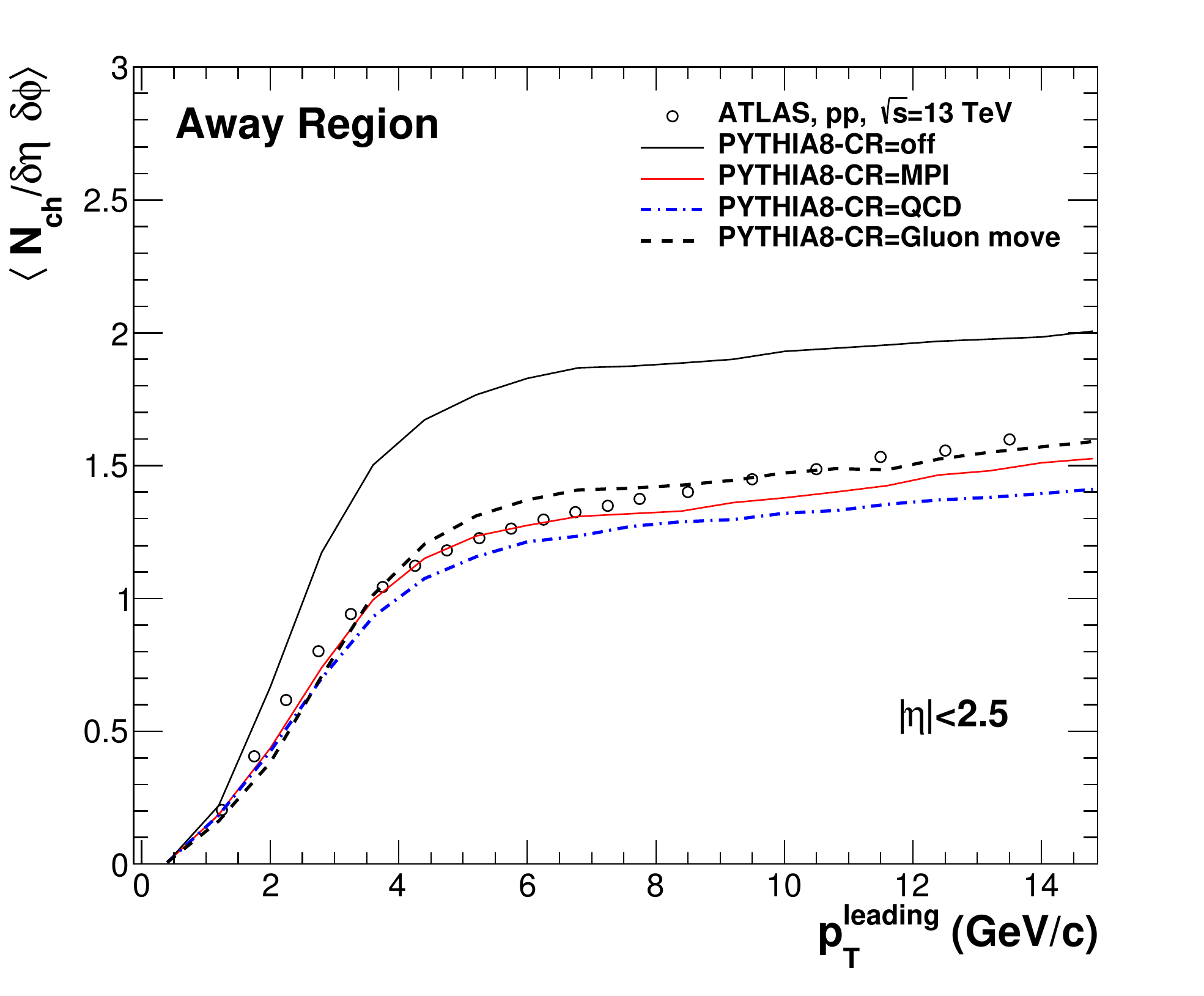}
\includegraphics[width=0.5\linewidth]{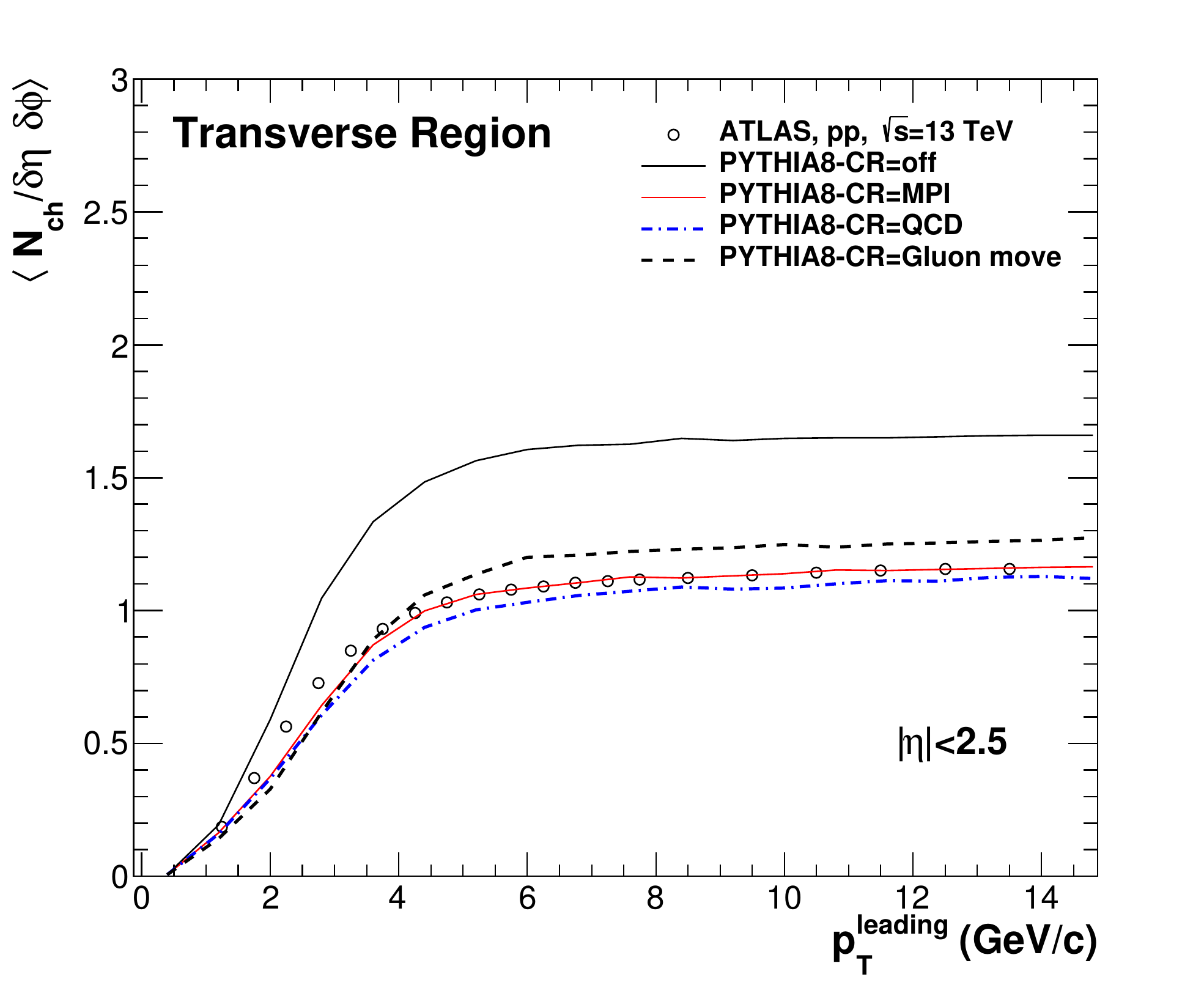}
\caption{ $\langle d^{2}N /d\eta d\phi \rangle$  as a function of  $p_T^{lead}$ for towards, away, and transverse regions at $\sqrt{s}$  = 13 TeV for different modes of color reconnections .}
\label{fig12}
\end{figure}

\begin{figure}
\includegraphics[width=0.5\linewidth]{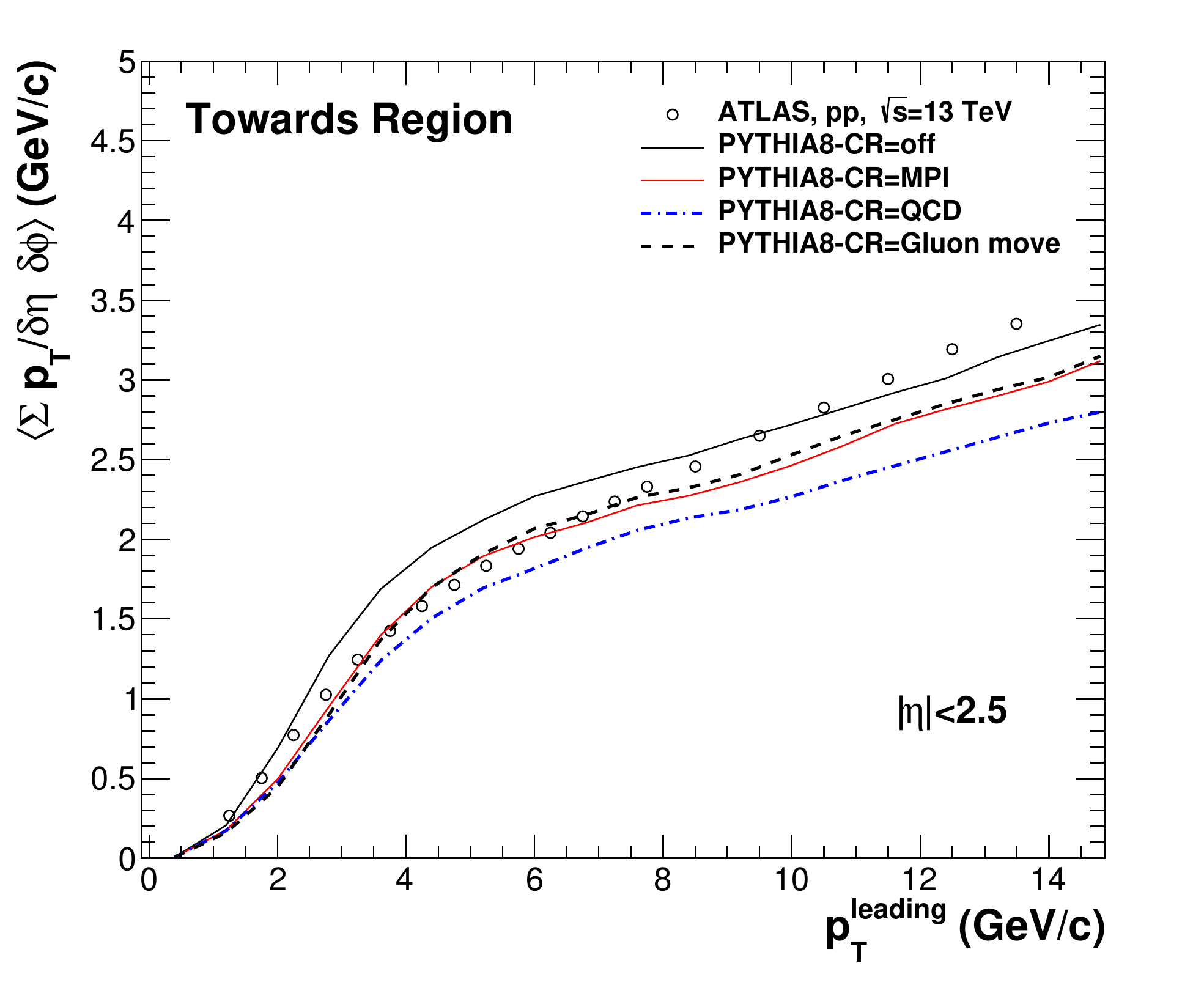}
\includegraphics[width=0.5\linewidth]{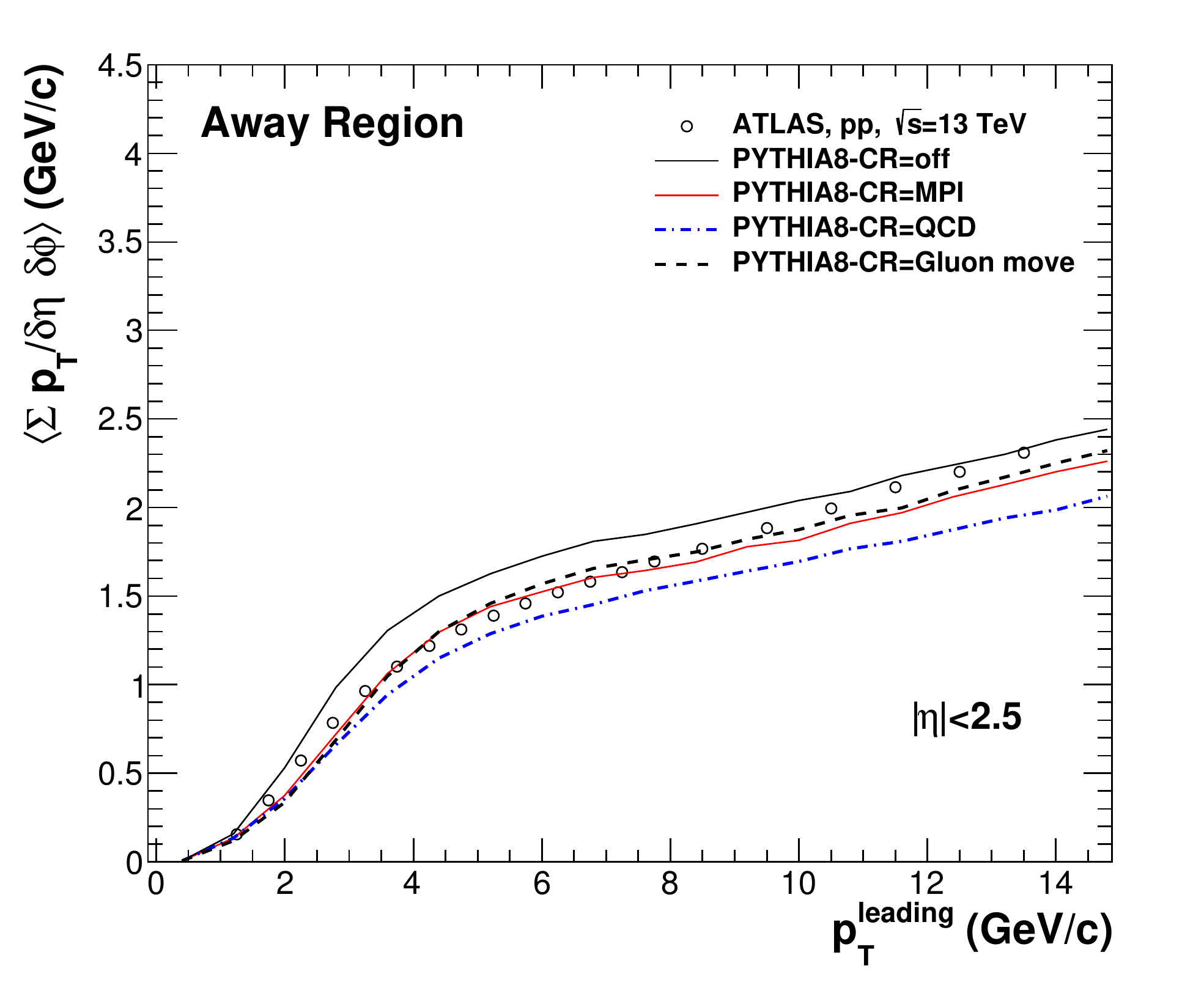}
\includegraphics[width=0.5\linewidth]{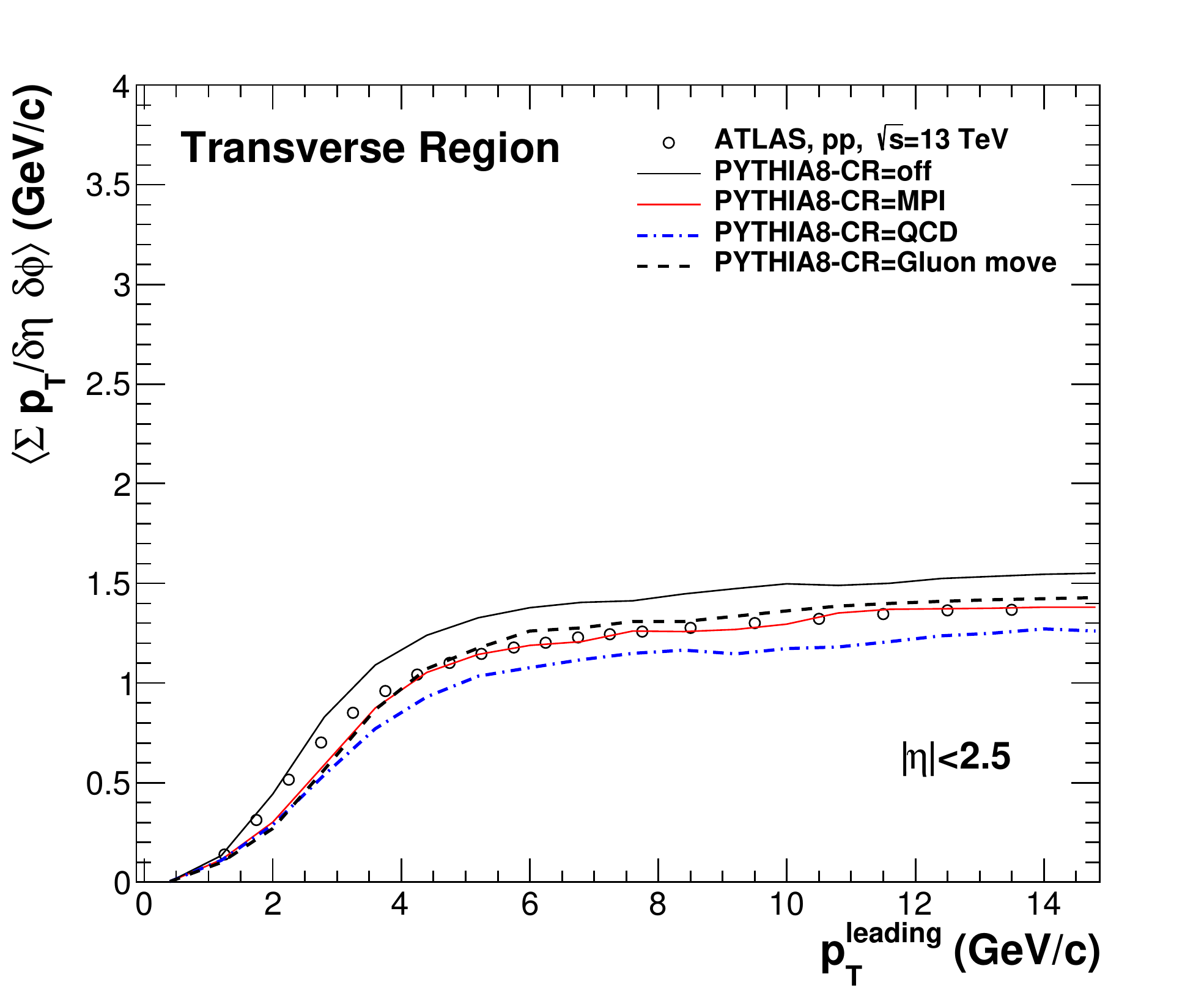}
\caption{ $\langle d^{2} \sum p_{T} /d\eta d\phi \rangle$  as a function of  $p_T^{lead}$  for towards, away, and transverse regions at $\sqrt{s}$  = 13 TeV for different modes of color reconnections .}
\label{fig13}
\end{figure}

The additional studies carried on effect of color reconnections can be  observed in  Figure \ref{fig12} and \ref{fig13}. The color reconnection mechanism plays a crucial role in the underlying events observables 
as it  governs the interactions between the partons of different $p_T$ scales before the hadronization process. It modifies the particle production in the events with a large number of multi-partonic interactions. It can be seen that the estimations with color reconnections are in good agreement with the measured data compared to the events without color reconnections which over-predict the data emphasizing on  the importance of color reconnection
mechanism \cite{ortiz2019color}. 
Furthermore, it shows the comparison of three different modes of color reconnection(CR ) : MPI-based, QCD-based and gluon-move based model implemented in Pythia 8. 

\begin{figure}
\includegraphics[width=0.5\linewidth]{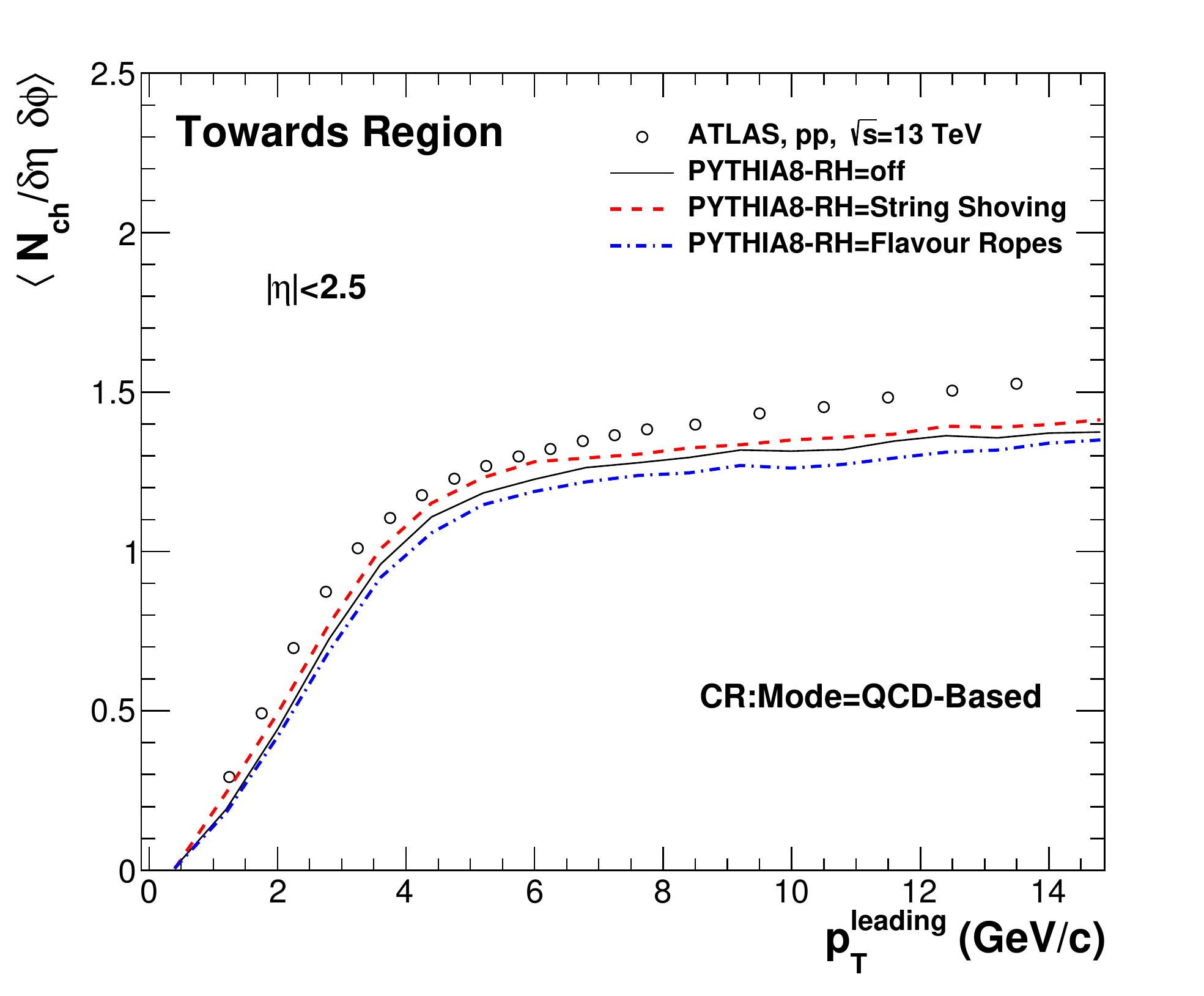}
\includegraphics[width=0.5\linewidth]{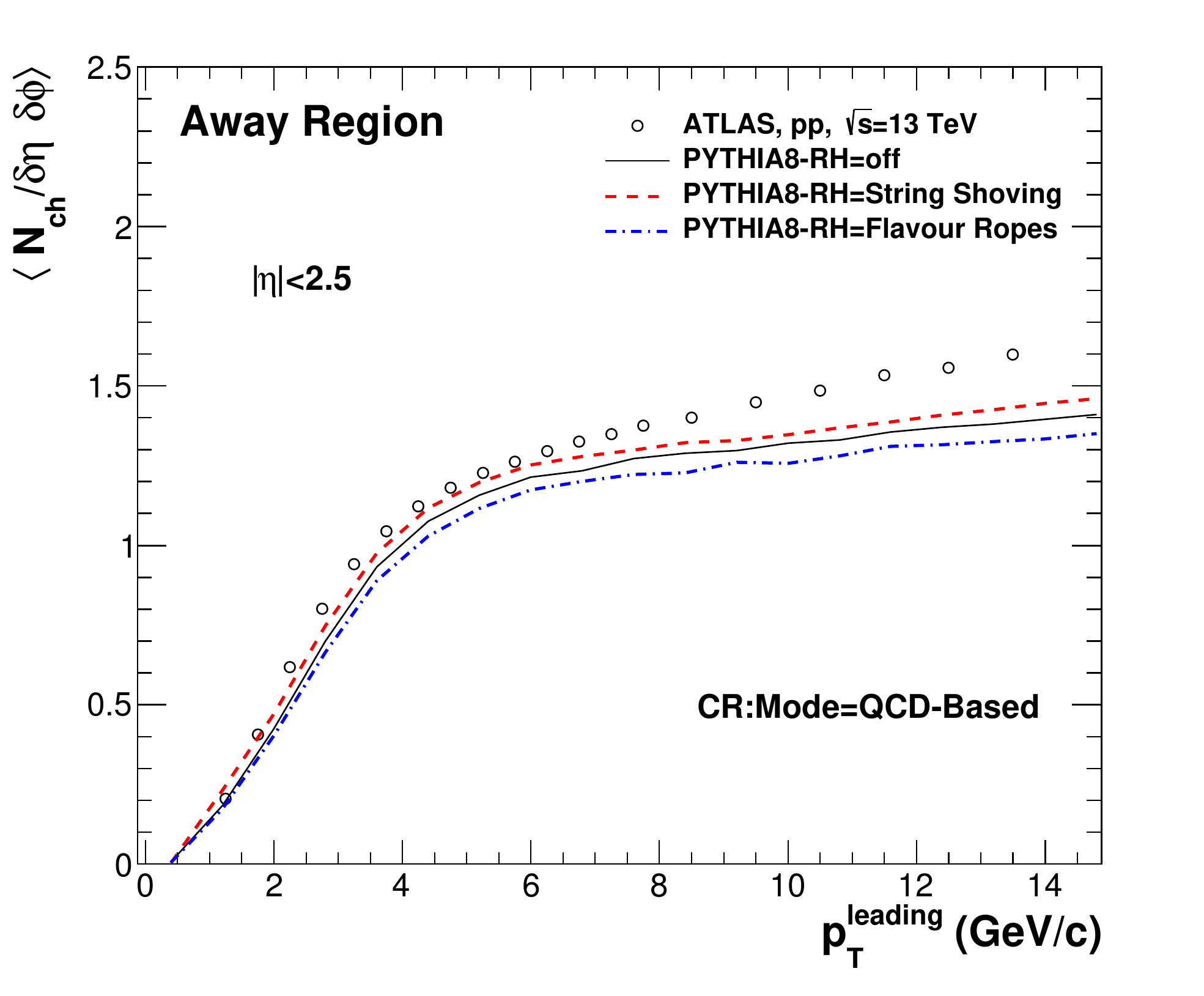}
\includegraphics[width=0.5\linewidth]{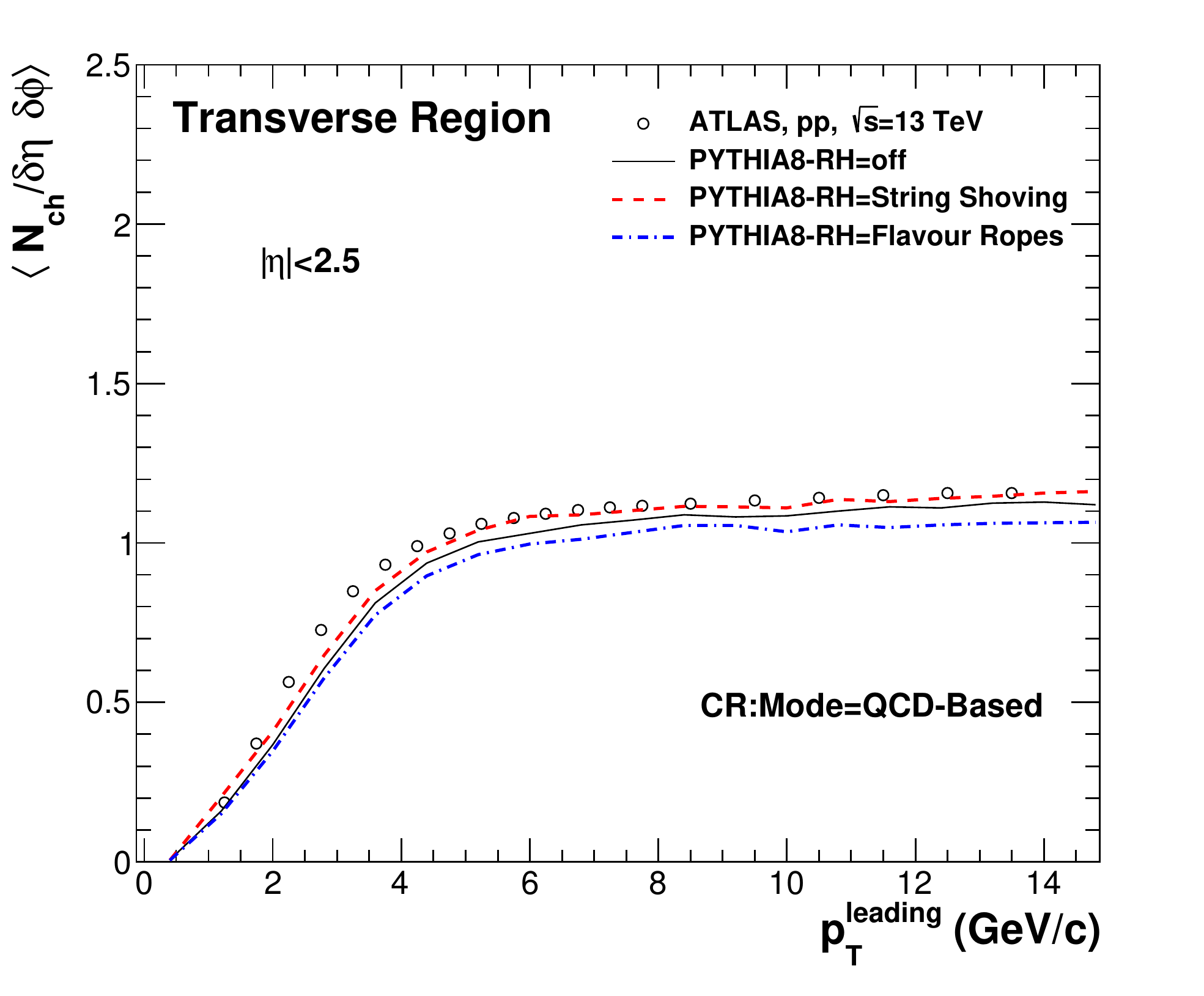}
\caption{$\langle d^{2}N /d\eta d\phi \rangle$ as a function of  $p_T^{lead}$  for rope hadronization with flavor ropes and string shoving.}
\label{fig14}
\end{figure}

\begin{figure}
\includegraphics[width=0.5\linewidth]{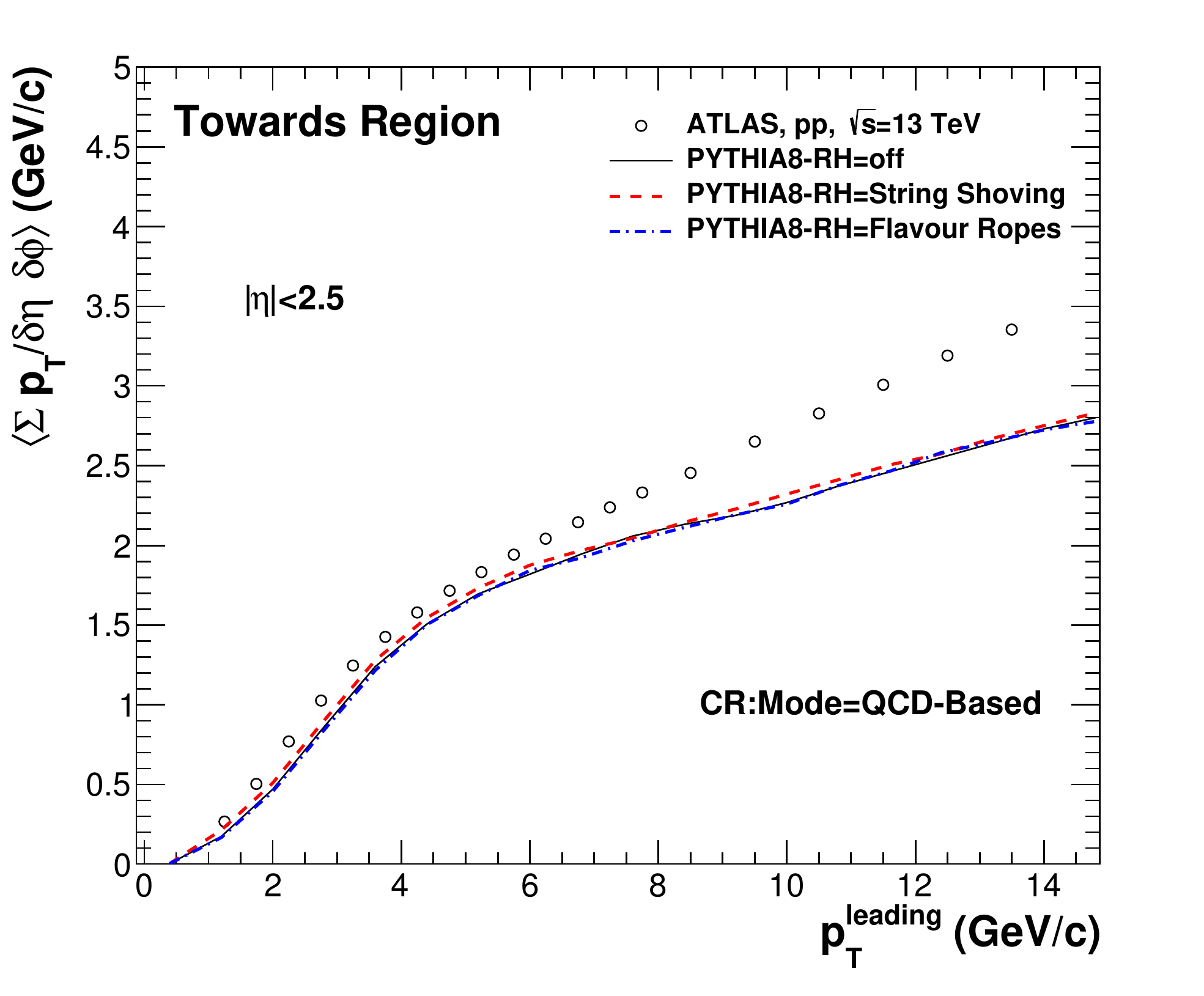}
\includegraphics[width=0.5\linewidth]{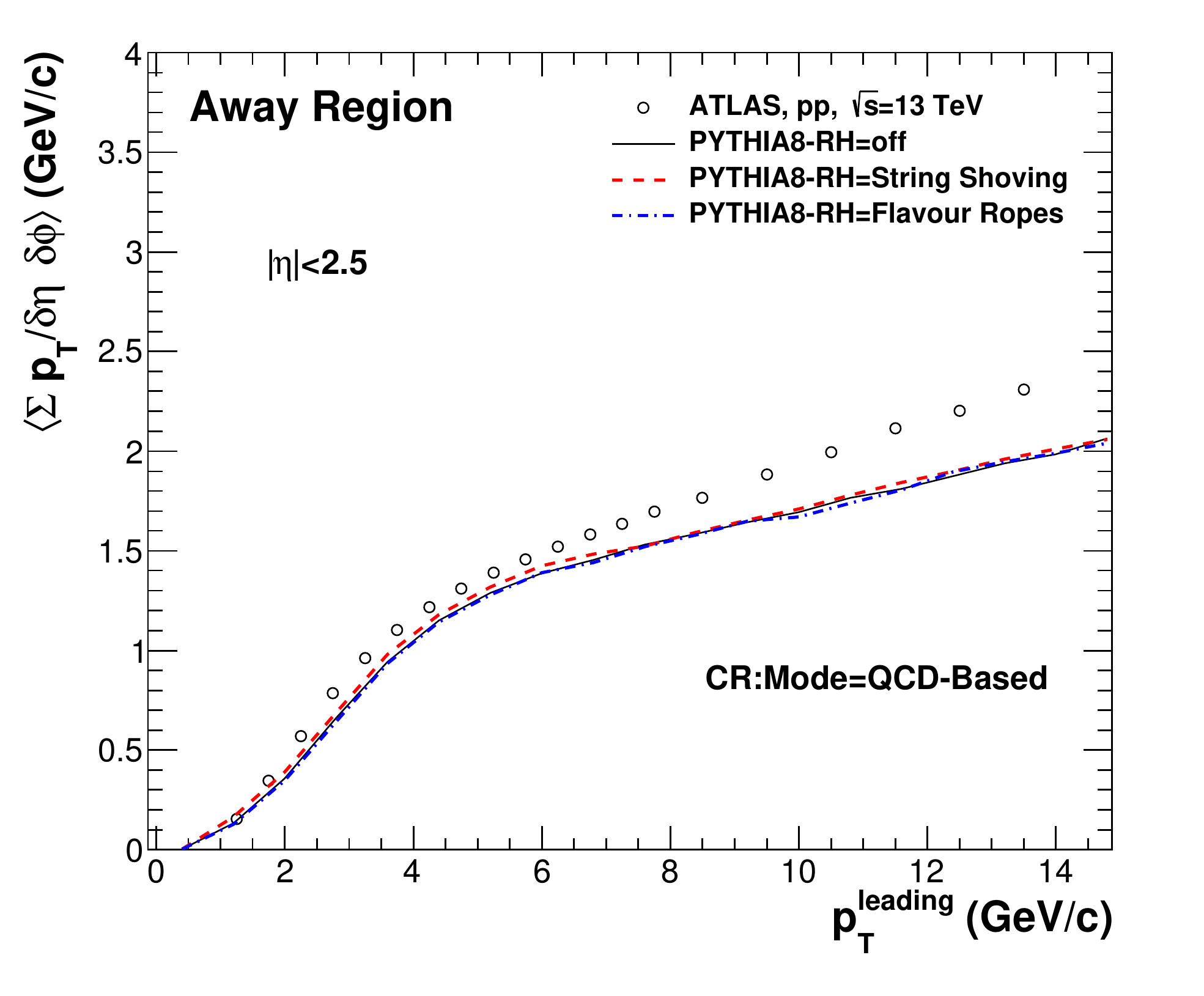}
\includegraphics[width=0.5\linewidth]{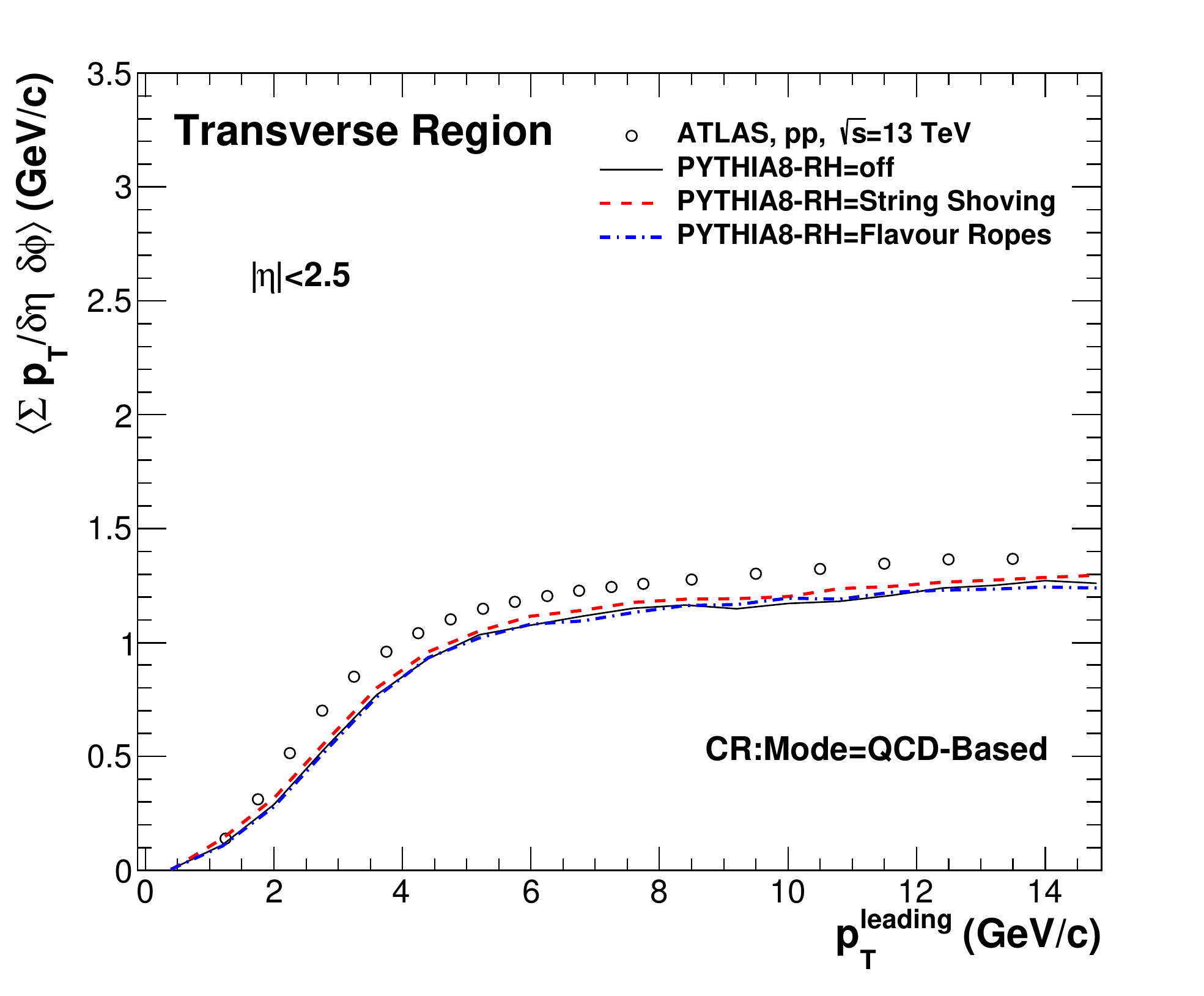}
\caption{ $\langle d^{2} \sum p_{T} /d\eta d\phi \rangle$  as a function of  $p_T^{lead}$  for rope hadronization with flavor ropes and string shoving.}
\label{fig15}
\end{figure}

The effect of rope hadronization with QCD-based color reconnection model  on  charged particle multiplicity density and mean scalar  sum $p_{T}$ has been shown as a function of  $p_{T}^{lead}$  at 
$\sqrt{s}$ = 13 TeV for towards, away, and transverse region in Figure \ref{fig14} and \ref{fig15} for the string shoving  and with flavor ropes . The rope hadronization mechanism describes the interaction of the overlapping strings produced in a small transverse area. The effect is more pronounced in high multiplicity events  and forms The string shoving mechanism refers to the pushing action of nearby strings before hadronization and can increase the multiplicity due to gluon excitations.  One observes an increase in mean charged particle multiplicity density for all three regions as an effect of the string shoving  and it describes the data better in the transverse region.  The switching of flavor rope mechanism enables the nearby strings  to  interact with each other to form color ropes. These ropes due to increased string tension preferably hadronize to  massive particles. Therefore, there is a slight decrease in multiplicity density.  There is  negligible effect of string shoving and flavour ropes on mean scalar $p_T$ sum.\\
These studies together with the predictions of other models like Herwig7 \cite{herwig1,herwig2} and Sherpa\cite{sherpa} can act as baseline to  understand the  processes at partonic and hadronic level to understand the contribution of underlying events in future experimental studies.  

\section{Summary}
The underlying event activities have been studied  in p$-$p collisions at $\sqrt{s}$ = 2.76, 7 and 13 TeV in  central and forward regions with  $|\eta| <$ 2.5 and -6.6 $< \eta <$ -5.2  respectively, using the Pythia 8 event generator. The hadronic activity due to underlying events has been studied by segmenting the azimuthal plane in towards, away, and transverse regions. The transverse region is sensitive to underlying event activity, while the towards and away regions receive dominant contributions from the primary hard scattering.  The observables, mean charged particle multiplicity density , $\langle d^{2}N /d\eta d\phi \rangle$ and mean scalar $p_T$ sum, $\langle d^{2} \sum p_{T} /d\eta d\phi \rangle$  were investigated as a function of leading $p_{T}^{lead}$. The effect of hadronic re-scattering,  different mode of color reconnections and rope hadronization has also been studied. This study can act as a baseline for fine tuning of event generators focusing on underlying events.

\section{Acknowledgements}
The authors would like to thank the Department of Science and Technology (DST), India for supporting the present work.

\end{document}